\RequirePackage{fix-cm}

\documentclass[twocolumn,natbib, enabledeprecatedfontcommands]{svjour3}          % twocolumn
\smartqed  % flush right qed marks, e.g. at end of proof
\usepackage{graphicx}
\usepackage[utf8x]{inputenc}			% Zeichenausgabe (vor fontenc!)
\usepackage[T1]{fontenc}				% Zeichenkodierung
\usepackage[english]{babel}				% Sprachunterstützung
\usepackage{microtype}					% Verbesserter Satzspiegel
\usepackage{amsmath}					% erweiterte Mathe-Umgebung
\usepackage{graphicx}					% Einbinden von Bildern
\usepackage{blindtext}					% Blindtext
\usepackage{booktabs}					% Tabellenlayout
\usepackage[counterclockwise]{rotating}
\usepackage{color}						% Farbdef. für farbige Links
\usepackage{nomencl}					% Nomenklatur
\usepackage{caption}					% Bildunterschriften
\usepackage{subcaption}					% Teilunterschriften
\usepackage{blindtext}					% Dummytext
\usepackage{amsmath}
\usepackage{bm}	
\usepackage{color}										
\usepackage{colortbl}
\usepackage{here} 

\usepackage{epstopdf}
\usepackage{multirow}

\usepackage[T1]{fontenc}
\usepackage{xcolor}
\usepackage{lmodern}
\usepackage{listings}
\usepackage{xurl}
\usepackage{graphicx}
\usepackage{float}
\usepackage{stfloats}
\definecolor{dunkelgrau}{rgb}{0.8,0.8,0.8}
\definecolor{hellgrau}{rgb}{0.95,0.95,0.95}
\usepackage{lineno}

\begin{document}
	\title{Struct-MRT: Immersive Learning and Teaching of Design and Verification in Structural Civil Engineering using Mixed Reality \\ \large State-of-the-Art Deployment Case Study for Steel and Concrete Structures}
    
	\titlerunning{Immersive Learning in Structural Civil Engineering via Mixed Reality}        
	\author{M.\,A. Kraus* \and I. \v{C}ustovi\'{c} \and W. Kaufmann} % 
	
\institute{
    *corresponding author: Dr. M.~A. Kraus, M.Sc.(hons) \\ 
    \\
    Dr.-Ing. Michael A. Kraus, M.Sc.(hons) \\
    Prof. Dr. Walter Kaufmann \at
	ETH Z\"urich\\ %
	Immersive Design Laboratory (IDL) / Design++ Initiative \\
	Institut f\"ur Baustatik und Konstruktion (IBK) \\
	Stefano-Franscini-Platz 5\\
	8093 Zürich \\
	\email{kraus@ibk.baug.ethz.ch}\\
	\\
	Preprint. Under review
}

\date{Version: 17.09.2021}

\maketitle
%\linenumbers

\begin{abstract}
Structural engineering lectures are core parts in undergraduate civil engineering curricula and the content demands advanced analytical thinking and abstract perception from real world structures to computational surrogate models and applicable design schemes. On the other hand, Extended Reality (XR) technologies such as Augmented, Virtual and Mixed Reality (AR, VR, MR) currently influence developments in many industries, including the AEC sector. Especially MR enables immersive experiences of blending interactive digital contents with the real. Our goal is to transform traditional paper-based instruction into an immersive lesson. To that end, this paper presents the conception, workflow and deployment of two MR applications for verification of typical yet geometrically complex structural members: a reinforced concrete corbel and a steel frame. The aim of this research is threefold: (i) to develop and implement the technological feasibility of such applications, (ii) to demonstrate possible use cases in the context of structural engineering lectures and (iii) to evaluate the presented MR examples and the future potential of such MR applications in structural engineering lectures through a survey. The workflow and especially the two MR teaching applications presented in this paper were developed based on Apple's ARKit and are thus natively deployable on iPhones and iPads. The verification process was reproduced in the MR applications based on conventional exercises that were taught on paper. Users can navigate independently through the applications and review every single step, including a true-to-scale, spatial representation of the specific component (concrete corbel or steel frame) as well as associated verification formulas in the respective step. Likewise, these applications were used to assess the demand and expectations for such immersive teaching techniques among students through a survey of 89 civil engineering students and instructors. The participants were asked to test the MR applications on their devices or watch pre-recorded video demonstrations, afterwards perception was elicited through a questionnaire. The results of a subsequent data analysis show a generally positive judgement of the MR application over the six questioned categories (style, usefulness, ease of use, enjoyment, attitude as well as intention towards using). The statistical analysis revealed some (positivity) biases for users with prior XR experience w.r.t. to usage and navigation, while inexperienced/novice users underlined increased enjoyment or excitement with this learning format. The outlook covers identified shortcomings and future developments in this field.

\keywords{Mixed Reality (MR), Engineering Education, Civil (Structural) Engineering, Interactive Learning Environments, Curriculum}
\end{abstract}

\section{Introduction}
\label{sec:intro}

A construction (e.g. building or bridge) comprises different information components, such as stresses, dimensions, materials, assembly, and construction. Furthermore the building life cycle is requesting different levels of detail and information: from conceptual design, schematic design, detailed design, construction documentation and fabrication towards assembly and as-built. Structural engineering plays a decisive role in all mentioned phases and has to cope with this great information complexity. Teaching structural engineering is hence a core part of education in every undergraduate civil engineering program and to some extent even in architecture and mechanical engineering. In these lectures, fundamentals of load bearing behavior and its structurally sound verification against state-of-the-art design standards or concepts for structures made of reinforced concrete, steel and timber are taught. The content in AEC lectures requires advanced analytical thinking and abstract perception, as real world (3D) structures under loading have to be modelled, i.e., transferred to computational surrogates for further material specific analysis. Especially for students of civil engineering with a lack of professional experience, this abstraction during lectures is a hurdle and thorough insight is gained by students in more depth only during follow-up exercises, tutorials or reflection after class.

It is especially alarming, that despite the crucial role of design and analysis in civil engineering curricula,and correspondingly comprehensive lectures dedicated to it, numerous studies \citep{Turkan2017,webster1996a,bowman2003a,rodrigues2008a} confirm a lack of sound comprehension of fundamental concepts such as load effects, load paths and the ability to visualize deformed shapes of simple structures by most students beyond theoretical formulae and methods. It is found in addition, that students struggle to relate basic structural members such as trusses, beams, and frames to more complex structural systems such as buildings and bridges and vice versa. One conclusion here is relating the deficiencies to the ineffectiveness of the traditional instructional techniques, which spend significant effort on investigation of discrete, isolated members with less emphasis on understanding the behavior of the entire structure in a three-dimensional context. The fact that this is not just a phenomenon of recent years is reflected in the assessment of \citep{ferguson1993engineering}, according to whom the focus in engineering education is too much on analytical methods and too little on visual, tangible perception.

Up to now, the design of individual components has mostly been demonstrated in frontal teaching. The students follow this process using paper based instructions, which sequentially specify the exact procedure for the verification. Thereby, traditional 2D orthographic projection drawings (i.e. elevations, sections and details) have been used in the vast majority of cases to convey information of an architectural or structural design idea to students. In some cases, realistic images of digital 3D models (renderings) were used during lectures or exercises to foster understanding of the content. While traditionally these renderings were drawn by hand, nowadays digital means of computer software such as Sketchup, Revit, AutoCAD, or ArchiCAD would be employed. Even if a digital model of a structure with its precise dimensions and defined materials etc. is handed to the student audience, perception is still two-dimensional through printing or computer screens. \citep{mckim1980experiences} captures the limited nature of discrete depictions in stating that "a graphic symbol is always less than what it represents" and "every graphic expression embodies a viewpoint, a single way of looking at reality" \citep{mckim1980experiences}.

Research indicates, that visual learning and use of visual aids are effective in conveying knowledge to students \citep{mcgrath2005visual,bransford2000people,fang2017quantitative}. Naturally, the way the structural elements to be dimensioned are spatially represented plays a significant role in the students' understanding. \citep{sorby2009developing} has analyzed the research on 3D spatial visualization skills of engineers and emphasizes their importance for professional success. In this context, work on real hands-on models is expedient for the development of such skills during studies. However, such models cannot be created for the large number of different problems in structural engineering due to economic and time constraints. Consequently, new, more practical ways must be found to enrich the visual representations and enhance students' understanding of the entire structure in a three-dimensional context.

Previous research showed the limited extent of using modern technology in structural engineering lectures to promote immersive and / or independent student learning. On the other hand, Extended Reality (XR) technologies have influenced (architectural) rendering in recent years due to successful developments and deployments of wearable devices and computer graphics, sparking attention and discussion of XR in education and practice within the AEC community, cf. Sec. \ref{sec:Literature_SoA_XR} and Sec. \ref{sec:Literature_SoA_XR_teaching}.

\begin{figure*} [h!]
	\centering
	\begin{tabular}{c}
	\hline  
	\cellcolor{hellgrau} \small Query history on XR methods in education over the past five years \\
	\hline  
	\vspace{.5mm}
	\includegraphics[width=0.9\linewidth]{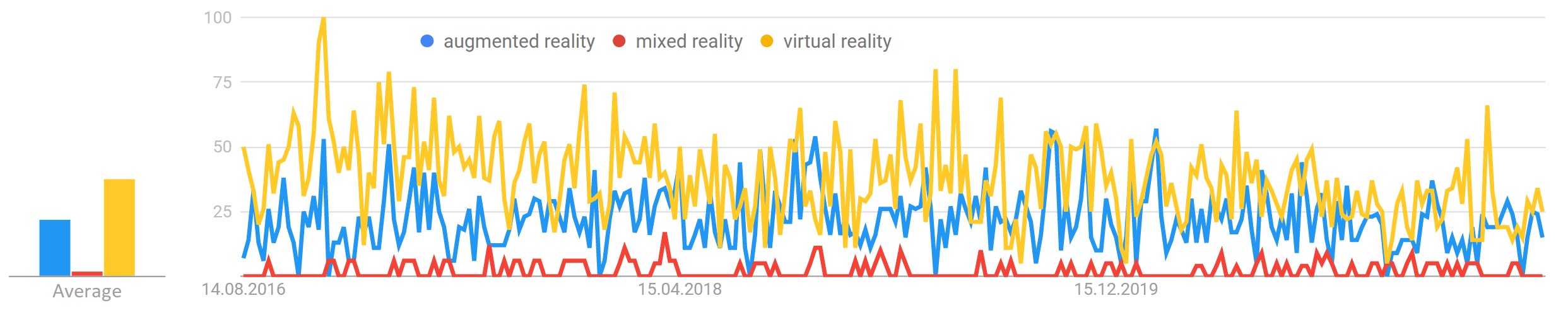} \\
	\hline  
	\end{tabular} 
	\caption{Results of a google trend analysis for XR methods in education: query history over the course of the last five years \citep{googletrendsSearch}}  
	\label{fig:googletrendsSearch1}
\end{figure*}

\begin{figure} [h!]
	\centering
	\begin{tabular}{c}
	\hline  
	\cellcolor{hellgrau} \small Global distribution of queries on XR methods in education \\
	\hline  
	\vspace{.5mm}	
	\includegraphics[width=0.97\linewidth]{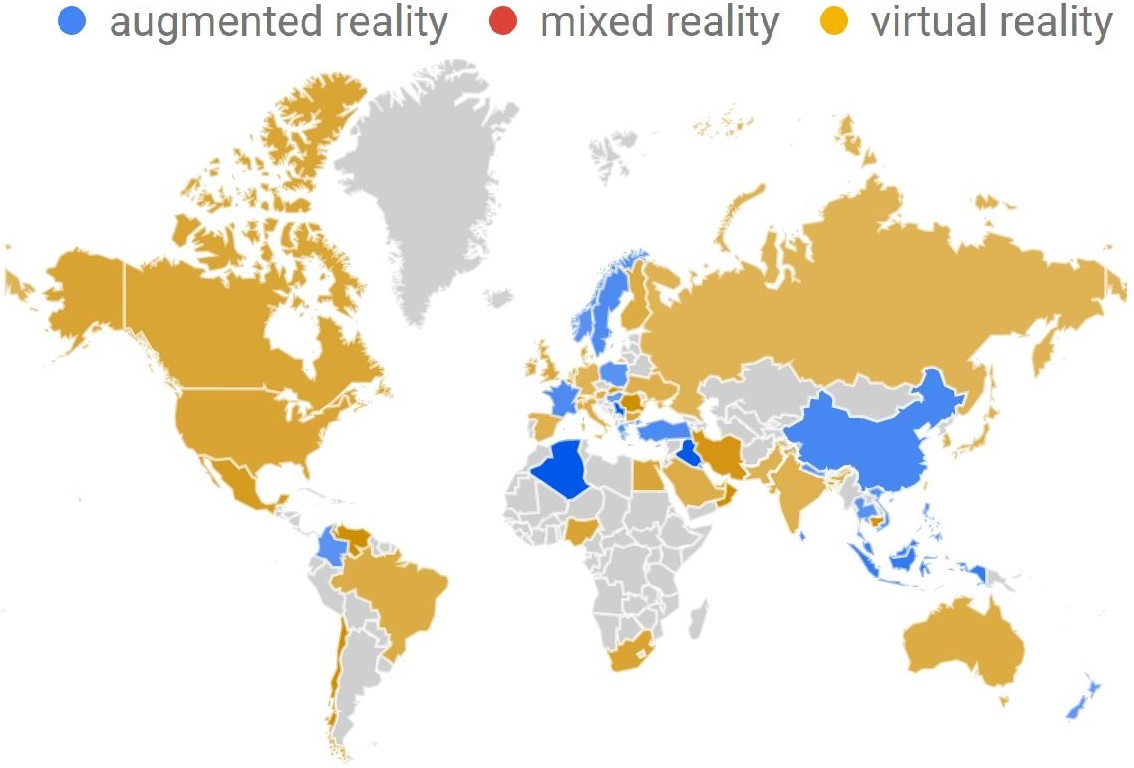} \\
	\hline
\end{tabular} 
	\caption{Results of a google trend analysis for XR methods in education: compared breakdown by region on global map \citep{googletrendsSearch}
 }  
	\label{fig:googletrendsSearch2}
\end{figure}

Therefore, before going into civil engineering specific XR applications, the authors analysed the results of Google trend data for queries concerning {virtual, mixed, augmented} reality (VR/MR/AR) in education over the course of the past five years on global scale. As shown in Fig.~\ref{fig:googletrendsSearch1}, there is a clearly visible hierarchy in people's technology awareness towards using VR in education, followed by AR and significantly less pronounced to MR. Another visualisation of the global query distribution according to the three XR categories and their use in education in Fig.~\ref{fig:googletrendsSearch2} shows, that different regions of the world emphasize different technologies on average. In the majority of countries, VR is the dominating subject of interest, whereas MR is not quite recognized all over the world. Based on the outline of "a lack of precise definition of MR" reported in \citep{speicher2019mixed} on the basis of expert interviews and a literature survey, the authors agree on the need for defining MR more thoroughly. To this end, the authors relate the reported numbers in Fig.~\ref{fig:googletrendsSearch1} mainly to the considered equivalence of XR technologies of most lay people. At this point, the further search for causes of these global differences remains subject to future research and reasoning.

In order to address the mentioned challenges in structural engineering education, the authors developed and implemented two Mixed Reality (MR) demonstrators for exemplary exercises from concrete as well as steel design and construction lectures. The main ideas are to: (i) enhance ordinary paper-based lessons with digital contents displayed via smart devices; (ii) use mobile technologies at student's disposal to support interactive, personalized and self-directed learning instead of {content/lecturer}-centered instructions; (iii) enhance student engagement and excitement about the lecture by immersion into the lecture content to foster deeper understanding at individual level; (iv) transform traditional instructional techniques in civil structural engineering towards a modern and digitized appearance. To this end, the authors conceptualized, implemented and deployed a mobile mixed reality tool (\emph{Struct-MRT}) for use in structural design and construction civil engineering courses. A complementary survey amongst students and teaching staff assessed its pedagogical potentials in engineering education.

The remainder of the paper is structured as follows: in Sec.~\ref{sec:Literature_SoA} some background on learning theories as well as XR is given. Sec.~\ref{sec:Methods} reports the methods employed in this study together with details. Sec.~\ref{sec:Development} describes the general conceptualization, implementation and deployment of the MR applications and specifies details for the two structural engineering design example use cases of a concrete corbel and a steel frame. Sec.~\ref{sec:Study} reports the data and results of the conducted survey about testing the MR applications. Sec.~\ref{sec:Conclusions} discusses the proposed approach, draws conclusions on the status-quo and the survey as well as for deployment and open research. Finally, Sec.~\ref{sec:Outlook} completes this with an outlook on future work and research.

\begin{figure} [b]
	\centering
	\begin{tabular}{c}
	\hline  
	\cellcolor{hellgrau} \small Cognitive structure and information processing model \\ 
	\hline  
	\vspace{.5mm}
	\includegraphics[width=0.95\linewidth]{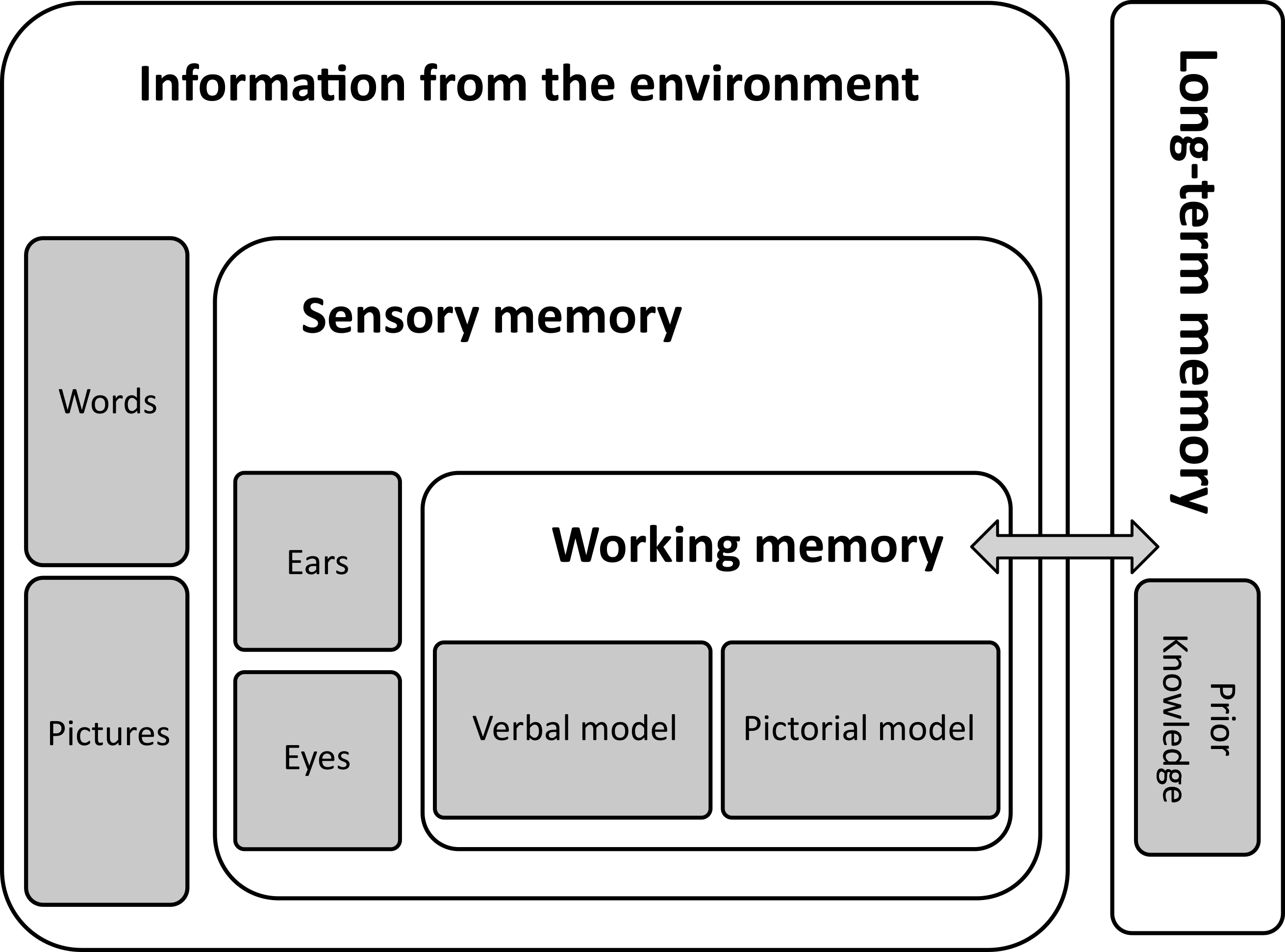}  \\
	\hline
\end{tabular} 
	\caption{Cognitive structure and information processing model, after \citep{shirazi2015design, ebbinghaus1964a}}  
	\label{fig:CogStructInfoProcess}
\end{figure}

\section{Background, Literature Review and State of the Art}
\label{sec:Literature_SoA}
This section serves to summarize essential background information on concerned topics such as learning theories, cognition and extended reality with a specific focus on civil engineering education.

\begin{table*}[]
\caption{Analysis of people's perception of real spaces and related cognitive processes. Compiled after \citep{ittelson1978a,carmona2003a,OsortoCarrasco2021}}
\label{tab:SpacePerceptionCognitiveProcesses}
\resizebox{\textwidth}{!}{%
\begin{tabular}{|l|l|l|l|l|}
\hline
  &
  \textbf{Process} &
  \textbf{Dimension} &
  \textbf{\begin{tabular}[c]{@{}l@{}}Type of obtained \\ Information\end{tabular}} &
  \textbf{Description} \\ \hline
\multirow{2}{*}{\textbf{1}} &
  \multirow{2}{*}{\begin{tabular}[c]{@{}l@{}} \textbf{Exploring}\\ \\ Individuals explore physical areas \\ for orientation, developing means \\ of locomotion and communication \\ within the given space\end{tabular}} &
  \multirow{2}{*}{\begin{tabular}[c]{@{}l@{}}\textbf{Cognitive Dimension}\\ \\ Individuals reason about, organize  \\ and store information in order to \\ make sense of the environment.\end{tabular}} &
  \begin{tabular}[c]{@{}l@{}}Determine the fixed or \\ given layout and \\ spatial boundaries\end{tabular} &
  \multirow{2}{*}{\begin{tabular}[c]{@{}l@{}}Does the user know about \\ activities and functions,  \\ which can be performed \\ in the environment?\end{tabular}} \\ \cline{4-4}
 &
   &
   &
  \begin{tabular}[c]{@{}l@{}}\\Environment experience\\ and definition\\\\\end{tabular} &
   \\ \hline
\multirow{2}{*}{\textbf{2}} &
  \multirow{2}{*}{\begin{tabular}[c]{@{}l@{}} \textbf{Categorizing}\\ \\ Individuals develop categories of \\  information or a taxonomy of\\ the environment by which\\ characteristics in a space are classified \\ in order to understand it.\end{tabular}} &
  \begin{tabular}[c]{@{}l@{}} \textbf{Affective Dimension}\\ \\ Involves feelings and how they \\ influence perception of the \\ environment and vice versa.\end{tabular} &
  \begin{tabular}[c]{@{}l@{}}Environmental Sensory\\ Information\end{tabular} &
  \begin{tabular}[c]{@{}l@{}} Does the user know if the \\ spaces in the environment \\ feel narrow, wide, low, \\ high, dark, or illuminated?\end{tabular} \\ \cline{3-5} 
 &
   &
  \begin{tabular}[c]{@{}l@{}}\textbf{Interpretative Dimension}\\ \\ The individual will rely on \\ memory points for comparison \\ with newly experienced stimuli.\end{tabular} &
  \begin{tabular}[c]{@{}l@{}}General and specific \\ information of the \\ environment\end{tabular} &
  \begin{tabular}[c]{@{}l@{}}Does the user know what \\ the things in the environ-\\ ment are made of? \\ Their materials, textures, \\ and size.\end{tabular} \\ \hline
\textbf{3} &
  \begin{tabular}[c]{@{}l@{}} \textbf{Systematizing}\\ \\ Individuals systematize the \\ environment through analysis \\ of environmental contingencies\\ (events happening in the environ-\\ ment).\end{tabular} &
  \begin{tabular}[c]{@{}l@{}} \textbf{Evaluative Dimension}\\ \\ It incorporates values and \\ preferences, and the deter-\\ mination of good and bad \\ elements in the environment.\end{tabular} &
  \begin{tabular}[c]{@{}l@{}}Environmental coherency \\ and symbolic meaning\end{tabular} &
  \begin{tabular}[c]{@{}l@{}}Is the user capable of desc-\\ ribing the environment in \\ a coherent way, explaining its \\ parts and different elements?\end{tabular} \\ \hline
\end{tabular}%
}
\end{table*}

\subsection{Supportive Learning Theories, Human Learning System and Constructivism}
\label{sec:Literature_SoA_Learning}
Psychology defines learning as change in knowledge due to experience \citep{schacter2009introducing}, which coincides with the definition of learning in artificial intelligence (AI) \citep{Goodfellow2016}. Literature reports several differences amongst deep learning and traditional classroom practices \citep{sawyer2006a, dewey1959a}, where the most relevant finding for this research is the result of a disconnection between class materials, students' prior knowledge and the perception of ideas not stemming from a textbook. \emph{Hence it is very important to understand the human information processing system before designing any new learning tool.} Acc. to \citep{feuer2002a} there are three fundamental principles (called the cognitive theories) of multimedia learning: (i) \emph{dual channels:} people have separate channels for processing verbal and visual materials, (ii) \emph{limited capacity:} people can process limited amounts of material in each channel in a given time, and (iii) \emph{active processing:} meaningful learning occurs when learners are engaged in appropriate cognitive processing during the learning process. The basic description of how the human information processing system works acc. to the cognitive theory of multimedia is shown in Fig.~\ref{fig:CogStructInfoProcess}. Following \citep{ebbinghaus1964a}, three memory storages can be distinguished: (i) \emph{sensory memory}: keeps information in the same sensory format as presented, has large capacity, but lasts only for a very brief time; (2) \emph{working memory}: holds information in an organized format, has limited capacity, and lasts for a short period of time; and (3) \emph{long-term memory}: stores information in an organized format, has large capacity, and lasts for long periods of time. In addition to multimedia learning, two other effects on learning have to be taken into account: social and cultural. Constructivism is one of the fundamental learning sciences based upon the two principles of (i) learners are active in constructing their own knowledge, and (ii) social interactions are important in the knowledge construction process \citep{bruning1999cognitive,Glasersfeld1996}. Acc. to \citep{mason2007a}, there are two flavours of constructivism. On the one hand, psychological/cognitive constructivism is defined as individually possessing knowledge, whereas on the other hand social constructivism understands learning as belonging to a group and participating in the social construction of knowledge. \citep{vygotski1978a} combined both concepts of psychological and social constructivism in the previously mentioned theory as social interaction coupled with cultural tools and activity shape individual development and learning.

\subsection{Environmental Perception and Recognition}
\label{sec:Literature_EnvironmentalPerception}
%Paper: Application of mixed reality for improving architectural design comprehension effectiveness
%
So far, the learning process and related theories were summarized. This section deals especially with the theories of environmental perception and recognition as this is a core feature of system and design comprehension resp. abstraction in structural engineering, cf. Sec.~\ref{sec:intro}. Perception of aspects and information about the design is complicated and non-obvious as the ideas and thoughts of the designer are encoded in the design and its \{geometric, alphanumeric\} means of deployment via 3D models or 2D plans. The receiver then tries to decode its meaning, based on prior knowledge and capability of understanding \citep{Yazdanfar2015}. In order to develop a supportive learning tool within this research, the understanding of the process of environmental investigation and perception with the aim of gathering an understanding (both virtual and real) is essential. \citep{lynch1960image} defines apparent clarity or legibility of a space as the ease with which its parts can be recognized and organized into a coherent pattern. Further authors \citep{carmona2003a,ittelson1978a} elaborate a definition of the process, which a person undergoes while trying to understand a physical space. This three stage process together with incorporated dimensions and kinds of obtained information is summarized in Tab.~\ref{tab:SpacePerceptionCognitiveProcesses}. The research presented in this study makes use of all three process steps and supports the users in different dimensions by providing information in an immersive way. To that end, previous research, which is concerned with immersive virtual environments through XR technology, found evidence, that XR provides a satisfactory representation of real physical environments \citep{pouya2013a,heydarian2015a,schmelter2009a} and XR representations are even perceived in the same way as real environments. Furthermore, immersive virtual environments allow an efficient and fast way of displaying or gathering essential information in an interactive style during the conceptual design phase, which is not possible with other technologies. In summary, XR technologies possess great potential for improving effectiveness of teaching design and verification processes in structural engineering by displaying environments and associated information in an intuitive way similar to real objects.

\subsection{Extended Reality (XR): Augmented, Virtual and Mixed Reality \{A;V;M\}R)}
\label{sec:Literature_SoA_XR}
\begin{figure}
	\centering
	\begin{tabular}{c}
	\hline  
	\cellcolor{hellgrau} \small Reality - Virtuality Continuum \\ 
	\hline  
	\vspace{.5mm}
	\includegraphics[width=0.99\linewidth]{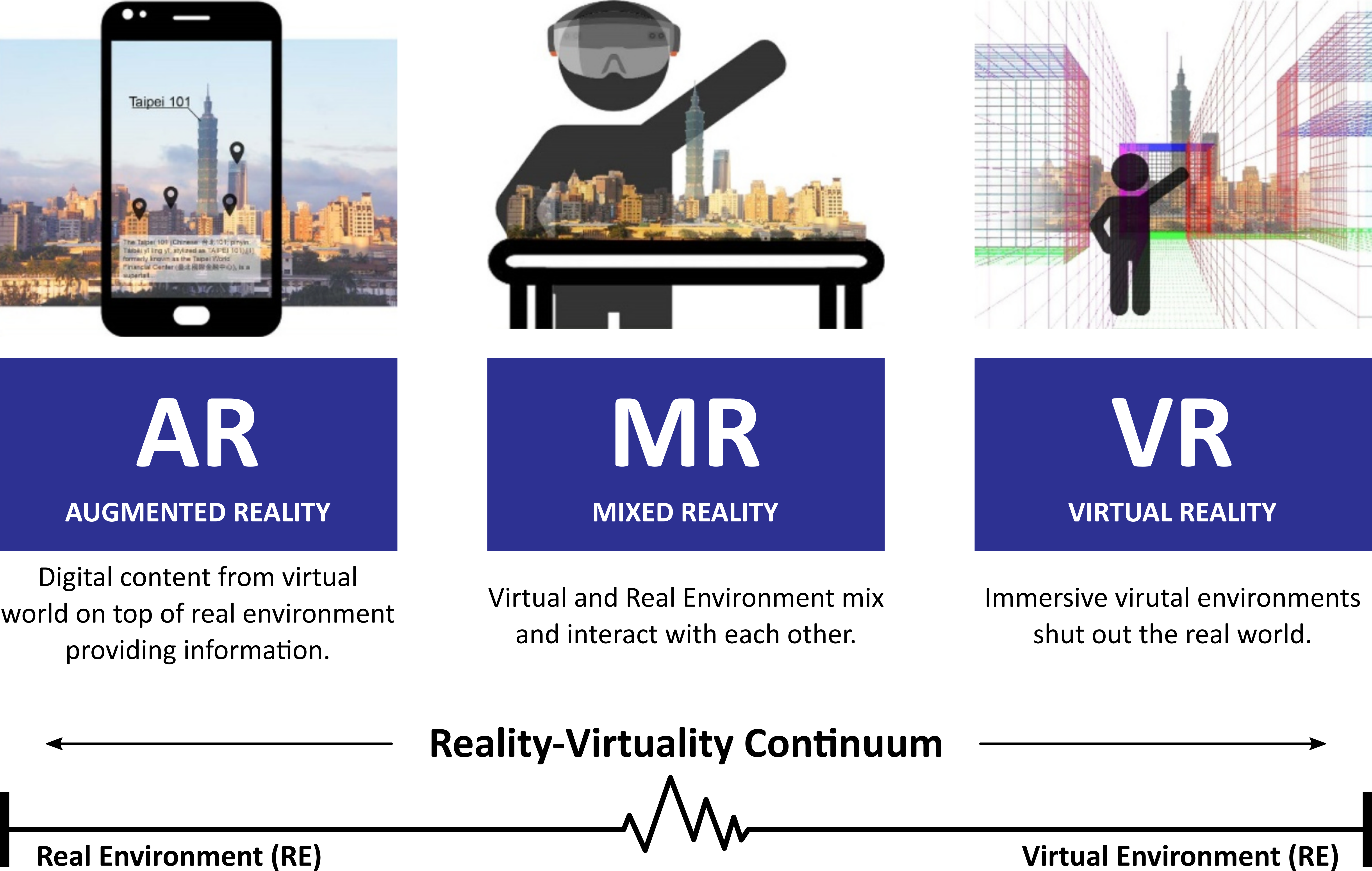}  \\
	\hline
\end{tabular} 
	\caption{Reality - Virtuality Continuum, after \citep{OsortoCarrasco2021,Milgram1999}}  
	\label{fig:VRARMR_explanation}
\end{figure}

The term \emph{Extended Reality (XR)} summarizes all technologies for combining real and virtual environments as well as human-machine interactions generated by computer technology and wearables \citep{fast2018testing,OsortoCarrasco2021,Dawood2020}. XR technologies create immersive digital worlds within the so called \emph{Reality-Virtuality Continuum} \citep{Milgram1999} to various extents depending on the intentions, cf. Fig.~\ref{fig:VRARMR_explanation}. Augmented Reality (AR) resides on the left hand of of the reality-virtuality continuum in Fig.~\ref{fig:VRARMR_explanation} where the real world is enhanced with digital content. As an example of AR, Fig.~\ref{fig:VRARMR_explanation} displays a smartphone camera pointing to a specific place or object and subsequently acquiring information about it on top of the displayed image. Further devices in that direction are "Google Glass" \citep{google2020a} head mounted displays (HMD) or "Bosch Smartglasses Light Drive BML500P" \citep{BoschSmartglasses}. Virtual Reality (VR) is located on the right hand side of the reality-virtuality continuum, where the user is immersed into a fully digital environment fading the real environment completely. An example device for this technology is Facebook's "Oculus Quest", which is already widely known and used within the videogame industry \citep{oculus_2021}. Mixed Reality (MR) hence is in-between the mentioned extremes and encompasses all technologies where computer generated content is blended in varying proportion with an individual's view of the real-world scene \citep{bray_2020}. MR thus produces new environments and visualizations with co-existence of real and virtual objects and a real-time interaction \citep{milgram1994a}. The currently most popular device in this category is the "Microsoft HoloLens" headset, where Microsoft states, that MR is a blend of the physical and digital worlds enabling humans, computers and the environment to interact \citep{bray_2020}. 

Decline in prices for most XR tools over the last decade allowed individuals to start exploring possibilities of XR devices in different fields for personal or professional purposes. Already back in 2015 smart glasses or head-mounted displays (HMD) gained broader public interest followed by numerouse VR HMD in 2016 \citep{hong201518}. Still, as per today, no devices exist, which can provide the user with an experience across the entire spectrum. Thus users must first clarify their main goal for the usage of XR and then choose a device accordingly \citep{OsortoCarrasco2021}. 

\subsection{Related Extended Reality in AEC Practice and Teaching}
\label{sec:Literature_SoA_XR_teaching}
% Mobile augmented reality for teaching structural analysis
XR methods (VR, AR) have been adopted to the AEC practice to notable extent for different phases of the lifecycle of a building so far. Earliest relevant applications of MR to AEC date back to the 1990s, where \citep{feiner1995architectural} applied MR in architectural design and urban planning using HMDs together with mobile computing technologies. Recent implementations of VR and AR are reported to support project planning, design, construction, and maintenance \citep{dunston2009key}; visualization of construction graphics \citep{behzadan2005a}; creation of virtual immersive sites \citep{xue2013a,henderson2009a}; construction defect management \citep{park2013a}; construction site visualization and communication \citep{woodward2010a}; progress tracking by comparing as-built and as-designed models, improving communication among project stakeholders, and planning and coordination for future projects \citep{kamat2010a,rankohi2013a}; excavator-collision avoidance systems \citep{talmaki2013a}; visualization of operations-level construction activities \citep{behzadan2009a}; and damage prevention and maintenance of underground utilities \citep{behzadan2009a,talmaki2014a}. The reported XR systems are either assembled as client-server systems on mobile phones \citep{woodward2010a}; as a combination of social media, XR, and 3D modeling \citep{siltanen2013a}; or as immersive XR models combining speech and gesture recognition for visualizing and interacting with buildings and their thermal environments interactively \citep{malkawi2004a}. \citep{Wang2006,Turkan2017,shirazi2015design, OsortoCarrasco2021} report a successful deployment of AR/MR methods with significantly less completion time, reduced orientation displacement, and lower task load for an AEC design request. Moreover, AR has recently been introduced to new application areas such as historical heritage reconstruction \citep{huang2009a}, training of operators of industrial processes \citep{schwald2003a}, system maintenance \citep{henderson2009a}, and tourist visits to museums and historic buildings \citep{wojciechowski2004a}.

Despite the mentioned existence of numerous studies together with the indicated promising potential of XR in AEC practice, there is only limited research on the development and integration of XR technologies into undergraduate AEC teaching, including the incorporation of gamification and cyber teaching tools \citep{yuan2002a,aparicio2007a,Kitipornchai2008,teng2004a,dinis2017virtual}.
%
% Application of mixed reality for improving architectural design comprehension effectiveness 
%
Still, evidence of XR facilitating learning of abstract and difficult-to-understand topics \citep{shirazi2015a,Sampaio2014} exists, showing that ambiguities in communication can be reduced or even eliminated using 3D visualizations as well as real-time interactions. Existing research proves lecturing not to be the best teaching approach as it fails to intrinsically motivate students and provides little, if any, incentive to build on existing knowledge \citep{felder2004a,felder2004b}. Especially engineering students join programs due to their interest in learning how to design and build buildings, towers and bridges \citep{matusovich2010a}. Current engineering education to a large extent fails to allow students to experience and understand the profession on a larger, application-based scale due to the limitations of traditional teaching methods and especially the historical disconnection of classroom and real-life practice together with a lack of hands-on possibilities \citep{Chen2014, Deshpande2011}. 
Instead of a holistic view of structures / components in a collaborative environment, the emphasis in teaching is placed on smaller-scale problems that can be solved by analytical methods \citep{ferguson1993engineering}. Meanwhile, the particular importance of 3D visualization capabilities for engineers, as highlighted by \citep{sorby2009developing}, remains unaddressed. XR will not only improve teaching and foster learning of structural analysis but rather change the instructional delivery, which has remained unchanged for a long time. \citep{Deshpande2011} discusses changes in the roles of lecturers and students made possible by XR methods and simulation games, where instructors act as promoters of knowledge, and students receive a more active and collaborative role. The urgent need for an introduction of XR methods in AEC teaching is underlined by the prediction of a broad XR industry adoption in the near future \citep{chi2013research} and the arising need for skilled and well-trained academics. XR integration into the AEC curricula hence prepares students better for the demands of the 21st century industry and the reported approach in this paper specifically enables them to exceed boundaries of traditional learning through immersive learning experiences and fostering the interest in the XR technology.

As this paper is specifically concerned with teaching structural engineering content, the literature review in the remainder of this section focuses especially on that field.
%Arezoo Shirazi: Design and Assessment of a Mobile Augmented
Several researchers have discussed that interactive simulations are more dominant for cognitive gain outcome across people and situations \citep{dede1998a,vye1998a,yoon2012a}. \citep{yuan2002a} developed web application for computer-aided learning of structural behavior. Construct3D is a 3D geometric construction tool designed for mathematics and geometry education \citep{kaufmann2003a}. \citep{Messner2003} implemented VR interfaces in undergraduate AEC programs allowing the students to develop a construction plan of a nuclear power plant within an hour with little experience in construction phases of this type of building. The authors found that immersive VR displays are beneficial for this type of lecture and the technology allowed an understanding for planning issues beyond their prior knowledge and visualization capacities concerning buildings and infrastructures. An educational AR application allowed users to interact with 3D content via web technology and AR/VR techniques \citep{liarokapis2004a}. \citep{Hafner2013} designed a practical VR course for university to teach students the use of VR hardware, software and applications for engineering topics. The study verified a higher motivation with the students at given tasks. Although AR is not a new technology it still has a significant potential for use in information delivery systems and more specifically in education \citep{behzadan2013a,billinghurst2002a}. An AR based system for enhancing spatial awareness and comprehension problems for engineering graphics education was designed \citep{chen2011application}. \citep{Yan2011} worked on an prototype for integrating Building Information Modeling (BIM) (providing geometric and alphanumeric information), and gaming to improve architectural visualization. \citep{pena2014a} researched a graphical application for tablet computers that supports 2D computer graphics and interaction for two-fold use: a teaching lab module and stand-alone instructional tool. \citep{shirazi2015design} designed and assessed the effectiveness of a collaborative context-aware mobile augmented reality tool (CAM-ART) for the civil engineering curriculum. \citep{Turkan2017} designed and developed an AR application and run it in a pilot study at a junior level structural analysis class to assess the pedagogical impact and the design concepts employed by the AR tool. Results of the pilot study indicated a positive effect of AR to students’ learning by providing interactive and 3D visualization features. Some works aimed at using simulation and multimedia as well as digital gaming to assist students in their understanding of the components and processes of building technology and sustainable design \citep{maldovan2005a,messner2005a,vassigh2003a,vassigh2008a,vassigh2010a}. \citep{dinis2017virtual} developed VR and AR applications for students of an introductory class of the Integrated Masters in Civil Engineering and tested those in two trials. Further successful development of VR applications in design and education tasks are reported by \citep{Sampaio2014,Wolfartsberger2019}. Relatively few researchers used AR-enhanced books and tabletop AR for student learning and training \citep{behzadan2013a,dong2013a} in construction and civil engineering so far.

\section{Research Methodologies}
\label{sec:Methods}
This research for developing a MR system for creating a workflow for teaching structural engineering lecture contents together with all experiments and the survey was conducted at the \emph{Immersive Design Lab (IDL)} at ETH Zurich in the first half of the year 2021 to shed light on this subject. During the mentioned time, the world faced a pandemic (COVID-19) causing many countries to alter the way and mode of working and learning. 

\subsection{Choosing XR for Lecturing Structural Engineering: VR, AR or MR?}
\label{sec:Literature_SoA_XR_choice}
%Paper: Application of mixed reality for improving architectural design comprehension effectiveness
% Mobile augmented reality for teaching structural analysis, Abs. 2.3
Summarizing the literature from the previous Sec.~\ref{sec:Literature_SoA_XR} resp. Sec.~\ref{sec:Literature_SoA_XR_teaching} and taking into consideration the results and recommendation, AR/MR technologies were chosen to further develop education support tools for structural civil engineering lectures due to safety and health reasons (being aware of the surrounding) together with the more natural immersive experience of AR/MR as the user remains in the "real world" while still inspecting and interacting with the 3D model without loss of the sense of location. AR allows a user interaction with the 3D models as if there were "real" physical models of the system under investigation placed on a table or in a room, enabling to walk around the 3D holographic object on site. This paper adopts several design concepts for the developed MR application: context-related 3D models, interaction and instant feedback, and collaboration. Literature from the previous section emphasizes the importance of relating 3D models to physical context and considers feedback as encouraging. In the majority of studies from Sec.~\ref{sec:Literature_SoA_XR_teaching} simple feedback is provided, e.g. an electrical circuit either works or does not. Necessary feedback in a MR application for teaching structural analysis on the other hand is way more complex as students are asked to determine forces or deflections and whether a structure meets design and performance requirements. This suggests that the MR application should be likewise ready to furnish relevant feedback with a similar degree of detail. Students are exposed to additional challenges in their perception of structural elements, as they may be used to 2D sketches of these structural members from textbooks yet MR uses a viewpoint projection. Further pros of AR und MR devices such as tablets, smartphones or the Microsoft HoloLens is that these are readily available, work without wired connections nor other attached devices, are thus wearable and incorporate sufficiently powerful CPUs to instantly generate high resolution content and allow for manipulating data in real time. They furthermore possess color displays and allow for intuitive use without much prior training to interact with the AR/MR content. 
% Mobile augmented reality for teaching structural analysis
Despite the fact that HMDs are viewed as typical AR display gadget, at present smartphones, tablets, and conventional PC displays are additionally utilized for AR because of their accessibility and minimal expense. Nonetheless, microelectronics and display technologies are persistently advanced further which thus brings down the expense of HMDs. An increased use of HMDs for AR applications can be anticipated in the nearer future and might replace smartphones and other mobile devices to a significant extend. Based on some prior experience of the authors with HMDs and the obtained technical and physiognomical challenges and in line with recommendations of \citep{Turkan2017, camba2014a,buesing2013a,singhal2012a,wu2013a,zainuddin2010a,billinghurst2012a} on similar teaching content, this study develops MR content for smartphones and tables, iPhones and iPads in particular. Smartphones and tablets are affordable and readily available amongst instructors and students, hence these kind of immersive lessons is highly scalable at low cost. Traditionally smartphones possess small screen sizes and hence appear less desirable than tablets for this application, but the recent trend to bigger screen sizes will facilitate an even better interaction with the MR content, which for tablets is already the case \citep{Turkan2017,camba2014a}. 

Given the XR technologies with their ability for creating and displaying virtual objects and virtual environments in an accessible and easy-to-use mode, the authors were motivated to develop \emph{Struct-MRT} and test its potential for use in real classroom situations and assess students' expectations towards improved learning through mobile devices allowing for an access to contextual visual information relevant to the course material. With that, the authors transform traditional paper-based instruction into immersive or \textit{mobilized lessons} (as defined by \citep{norris2008getting}). The objective thus is to move away from a content- / teacher-centered teaching style to a systematic student-centered approach that allows for individualized and self-directed learning \citep{looi2009anatomy}. Finally it should be noted, that the authors did not see the need to look at various MR application designs since various visual features and collaboration concepts were researched in past examinations from several domains. This study rather adopts "best practices" from literature to verify usefulness in practice.

\subsection{Mixed Reality Application Development}
To elaborate on the perceived opportunities and challenges associated with the design and deployment of a MR teaching application for structural engineering lectures, the methodology of collaborative brainstorming with a follow-up was chosen \citep{drummer2011forschungsherausforderung}. This was conducted in five steps:
\begin{itemize}
    \item preparation of a pre-structured mind-map in a brainstorming tool by the speakers of the author team.
    \item collaborative brainstorming with evaluation and discussion of opportunities and challenges (items).
    \item selection (and reformulation, if necessary) of the primary opportunities and challenges.
    \item sorting out items that are less MR specific (eg. legal hurdles in technology use in general).
    \item literature review on the research status of each item.
\end{itemize}

Based on the results of the previous work steps, the MR development process followed in this research included the following steps:
\begin{itemize}
\item Identification and evaluation of existing software for data exchange between CAD and MR
\item Deduction of a Workflow for Design of Lectures using MR
\item Conducting first "practice tests" with two developed MR applications (concrete corbel and steel frame) with subsequent survey among students and lecturers at the department of Civil Engineering at ETH Zurich
\item Derive the research and development needs regarding software, interfaces and MR end-devices for the use of MR in civil engineering education.
\end{itemize}

\subsection{Survey}
Within this research the survey method was used to gather information about student's perception and rating of the developed two MR teaching demonstrators as well as to derive future potentials. The survey within this study was adapted from \citep{Turkan2017} and enhanced by additional questions. More detailed information on the survey can be found in Sec.~\ref{sec:Study_construction}.

\begin{figure*} [h!]
	\centering
	\begin{tabular}{c c c c}
	\hline  
	\cellcolor{hellgrau} \small (a) MR App Concrete Corbel & \cellcolor{hellgrau} \small (b) MR App Steel Frame  & \cellcolor{hellgrau} \small (c) Video MR Concrete Corbel & \cellcolor{hellgrau} \small (d) Video MR Steel Frame \\
	\hline  
	\vspace{.5mm}
	\includegraphics[width=0.22\linewidth]{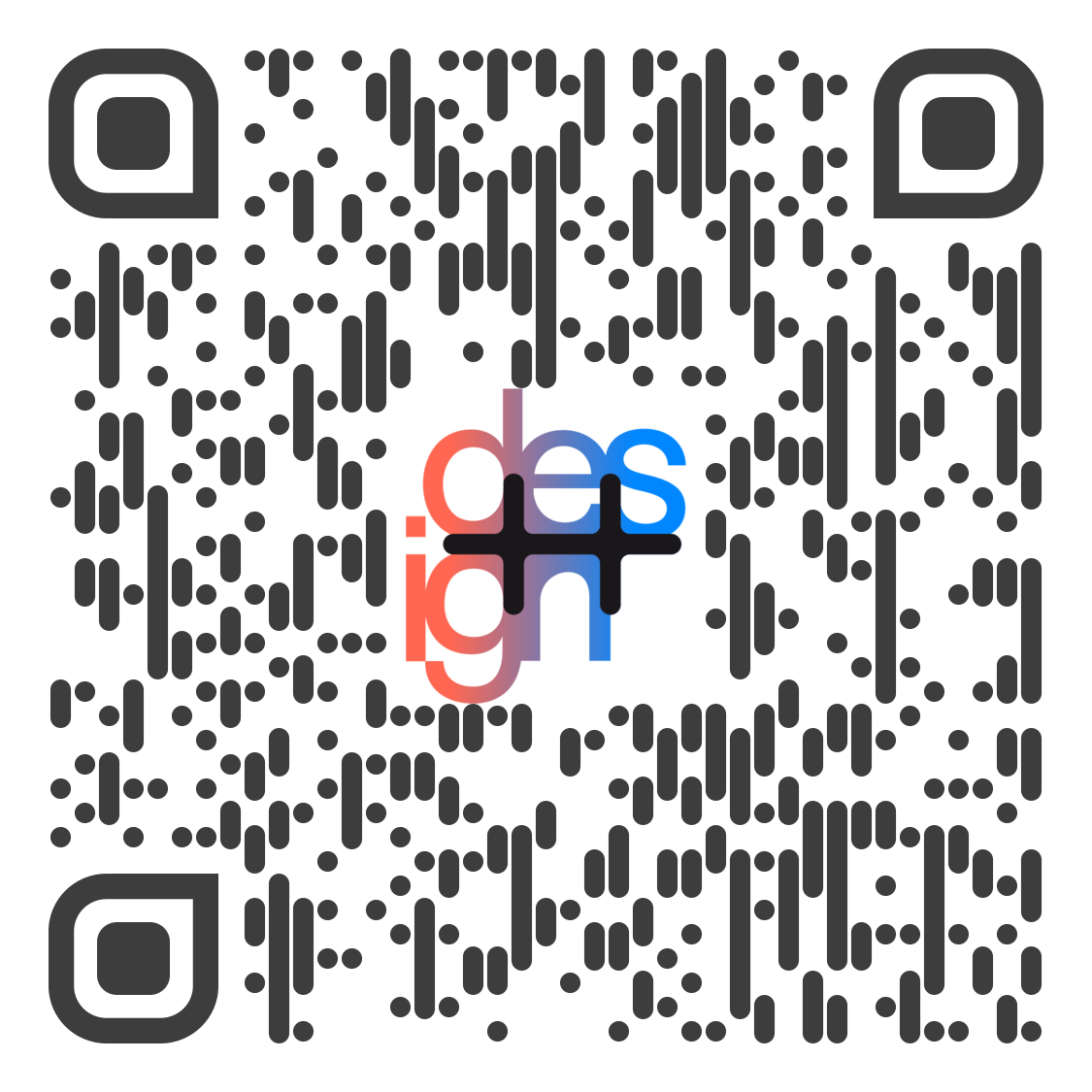} & \includegraphics[width=0.22\linewidth]{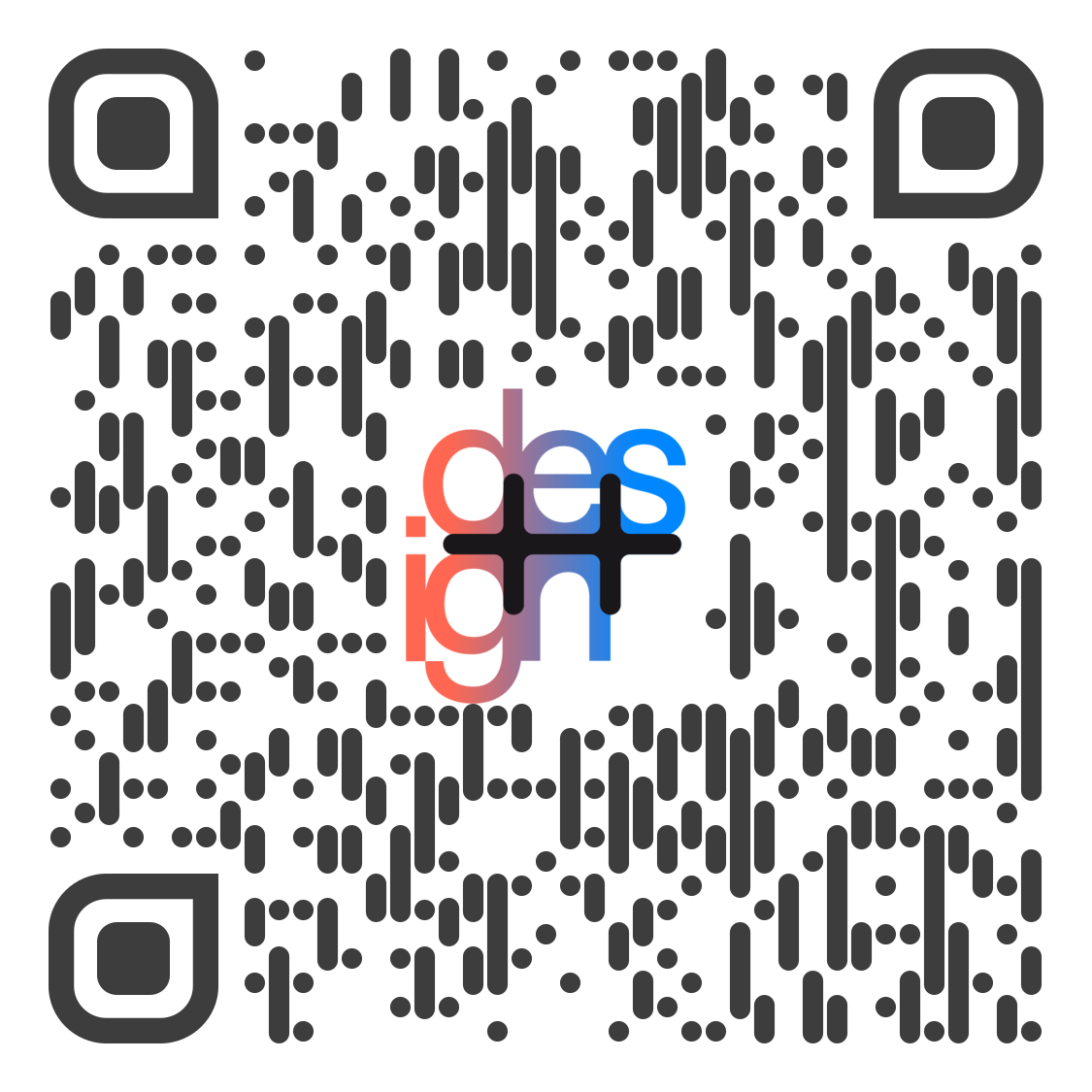} & \includegraphics[width=0.22\linewidth]{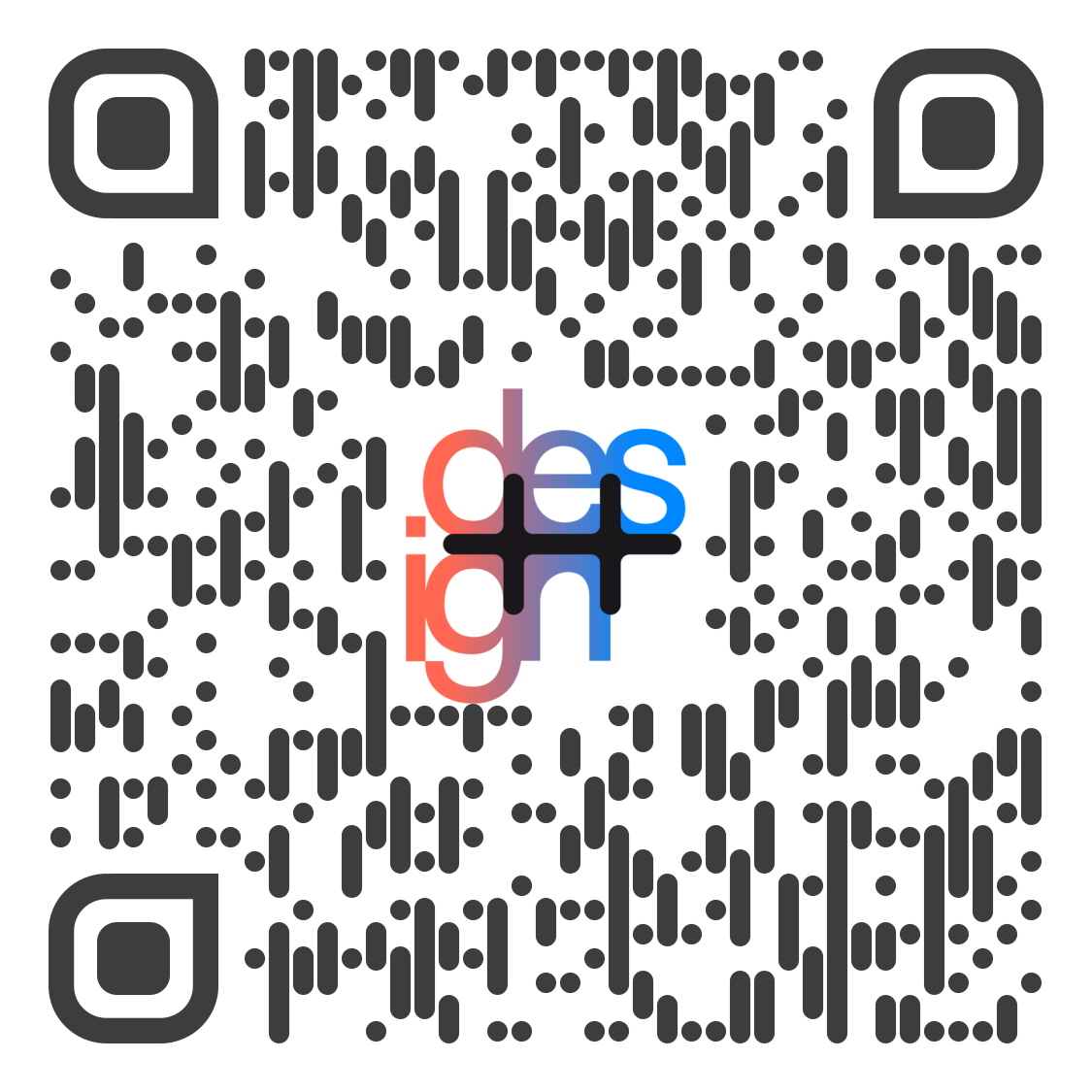} & \includegraphics[width=0.22\linewidth]{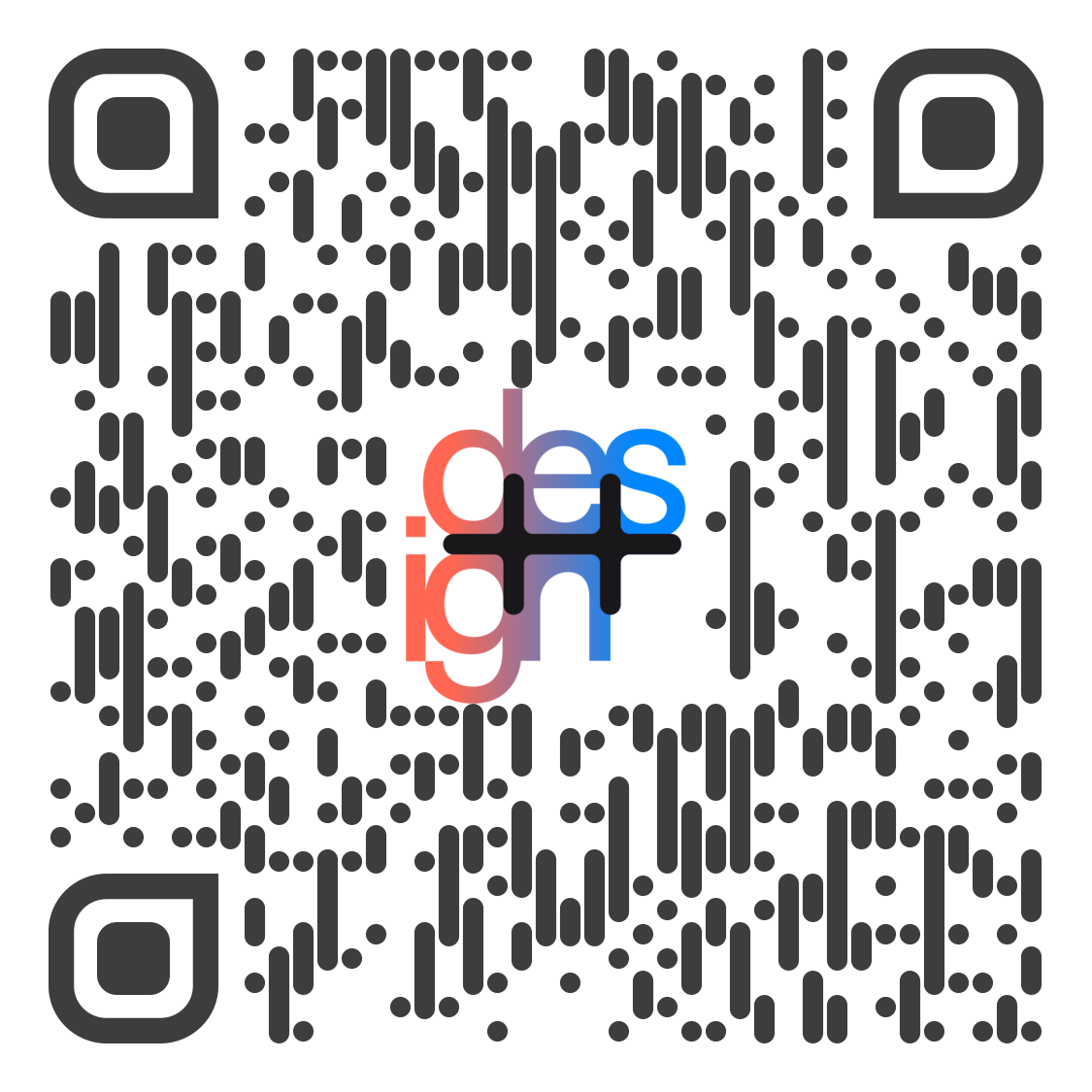} \\
	\hline
\end{tabular} 
	\caption{QR Codes of the (a) MR App of Concrete Corbel; (b) MR App Steel Frame; (c) Video of MR Concrete Corbel; (s) Video of MR Steel Frame; }  
	\label{fig:QRCodes_AppsVids}
\end{figure*}

\section{Development of Mixed Reality Applications for Structural Design and Verification Lectures}
\label{sec:Development}
This section presents a self-developed MR application designed for iOS-based systems (iPhones / iPads) to illustrate the complex teaching content of structural engineering such as the static modeling of systems and details as well as their consecutive dimensioning together with constructive, standard-compliant implementation in an immersive, playful manner. These applications allow the user to interactively navigate through the verification of a structural component. In addition to the true-to-scale, three-dimensional representation of the structural components, supplementary texts and formulas are displayed dynamically to explain the verification process. The workflow can easily be transferred to other operating systems (e.g. Android) without impacting the educational findings of this paper. ARKit was used to create two demonstrator applications, covering the contents of two sample exercises for the design of structural components taught conventionally on paper: (A) concrete corbel and (B) steel frame (visual impressions can be found in Figures \ref{ScreenshotARebar} and \ref{ScreenshotBFrame}).

Application A deals with a reinforced concrete corbel that requires the students' spatial imagination skills especially for the design and arrangement of individual reinforcing bars. The verification procedure presented in MR consists of the geometry of the component, the acting loads, the underlying strut-and-tie models and the final reinforcement layout. This rather simple example is meant to illustrate how the development process of an MR application can look like for a specific civil engineering problem. The more extensive application B deals with a steel frame for which the structural details of the apex joint, frame corners and column bases are to be designed. The challenge for student comprehension in this structure is that the details consist of many individual parts and the contribution of each component to the load-carrying behaviour must be understood in order to perform a thorough design. This application includes both the MR representation of the entire frame with global dimensions and acting loads, as well as separate, detailed visualizations of the individual structural details, accompanied by design formulas. The objective here was to investigate how the sequence of a real dimensioning from the global system to the verification of individual components can be reproduced as accurately as possible in an MR application. In both examples, the individual 3D objects as well as the associated verification formulas are shown in an MR environment and can be studied step by step by the students. Emphasis was placed on giving students a more comprehensive view of the structure rather than just 2D drawings, while dynamically providing only the information at each step that is valuable to them for design. Another focus of this study was identification and evaluation of existing software for data exchange between CAD and MR. This section introduces the application design and the software architecture. In addition to the technical feasibility, the practical application in teaching was also investigated as part of this study in Sec.~\ref{sec:Study}. 
%
%Therefore the MR applications were used for a self-guided hands-on demonstration followed by a survey among students and teaching staff. 
%
This was intended to capture attitudes towards the use of MR for civil engineering education as well as to derive the research and development needs regarding software, interfaces and MR end devices.

The MR demonstrators (currently implemented for Apple iOS and iPadOS devices only) and demonstration videos can be gathered via the QR codes as provided in Fig.~\ref{fig:QRCodes_AppsVids} or alternatively via \url{https://polybox.ethz.ch/index.php/s/hOH1VXBufmlrcpM} (demonstrators) and \url{https://polybox.ethz.ch/index.php/s/ZdcTEZnM6YiAaf7} (videos), respectively. 

\subsection{Software Architecture}
\label{subsec:SoftwareArchitecture}

\begin{figure}
	\centering
	\begin{tabular}{c}
	\hline  
	\cellcolor{hellgrau} \small ARKit and RealityKit Overview \\ 
	\hline  
	\vspace{.5mm}
	\includegraphics[width=0.95\linewidth]{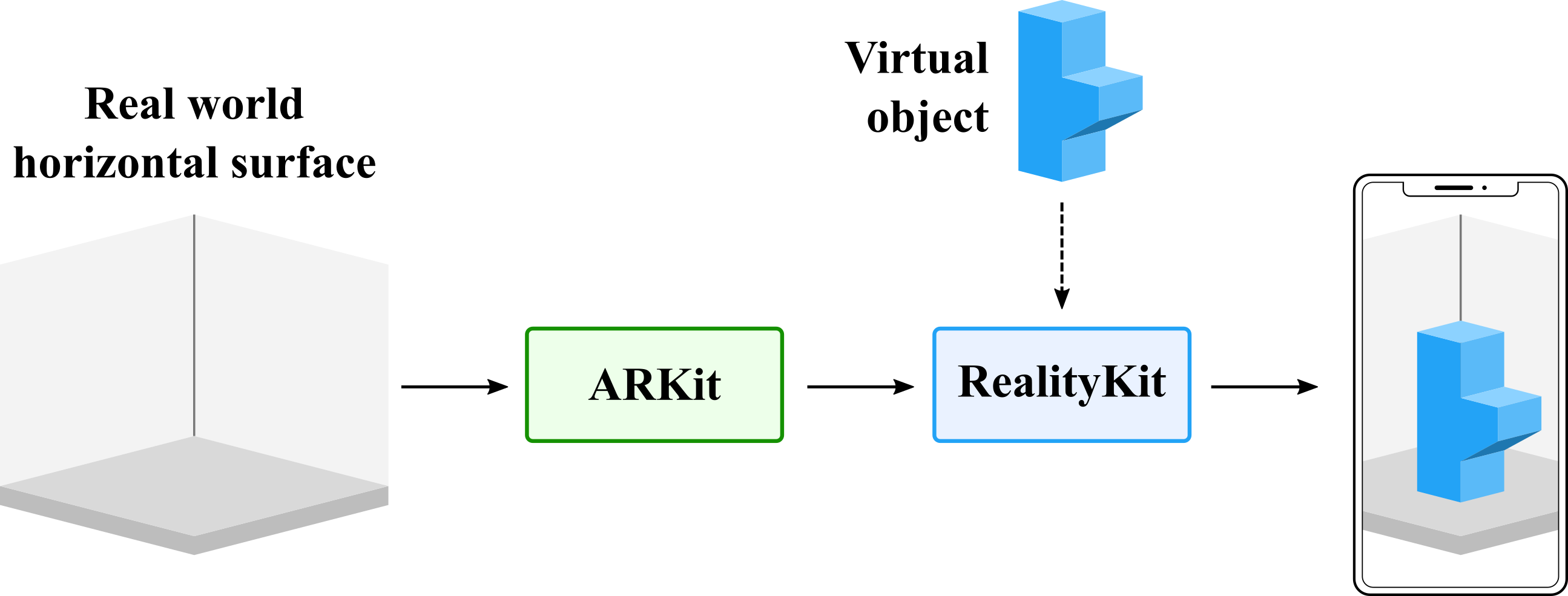}  \\
	\hline
\end{tabular} 
	\caption{Overview of ARKit and RealityKit, after \citep{appleRealityKit}}  
	\label{fig:ARKit_World}
\end{figure}

In developing and deploying of \emph{Struct-MRT} on iOS devices, the Apple \textit{ARKit} \citep{appleARKit} served as a technical framework for integrating motion features and camera modules as well as enabling native display of AR / MR content without need for installing any additional apps (software requirements iOS 11 or higher). \textit{ARKit} combines the device’s sensors inputs with advanced scene processing to create an AR experience. Visual-Inertial Odometry (VIO) is used to accurately update the position of the device and other objects and are thereby mapping the surrounding environment in real-time. ARKit uses \textit{ARSession} to manage the AR experience, e.g. managing the input from the hardware (camera and motion sensors) and performing an analysis on captured images. \textit{ARSession} then interprets the data stream for establishing a relation between real and virtual environment (created by \textit{ARKit}) allowing for adjustments to changes in the environment in real-time. \textit{ARSession} utilizes the  \textit{ARView} renderer for interaction with the current session and customized functions for performing various tasks, such as object placement in the real world. This is achieved through the \textit{ARView} Scene instance, where only one instance of the scene object exists for the session and it acts like a container that keeps track of the entities rendered by \textit{ARKit}. A configuration is required to run an \textit{ARSession}, where multiple options exist in dependence of the main task of the application. As the purpose of the applications of this paper is mainly tracking the device position and mapping the environment around it, the used configuration is \textit{ARWorldTrackingConfiguration}. By this configuration, applications are enabled to augment the environment and track position of the device relative to any object. The configuration is tracking the movement in a 3D space by using six degrees of freedom (6DOF): rotation (roll, pitch, and yaw) and position (x, y, z). The created world tracking session can be configured to recognize and interact with objects in the real world that is captured by the camera system and has numerous more viable features. In this paper, \textit{RealityKit} is used, leveraging data from \textit{ARKit} to place virtual content on detected surfaces in the spatial view on end devices together with correct scaling and shading. Apple's software \textit{Reality Composer} provided a graphical interface for assembling, testing and publishing of MR compositions. An overview of the relations between the modules discussed in the latter of this section is given in Fig.~\ref{fig:ARKit_World}.

\begin{figure*} [h!]
	\centering
	\begin{tabular}{c}
	\hline  
	\cellcolor{hellgrau} \small Process for developing and deploying MR applications based on 3D CAD Models \\ 
	\hline  
	\vspace{.5mm}
	\includegraphics[width=0.95\linewidth]{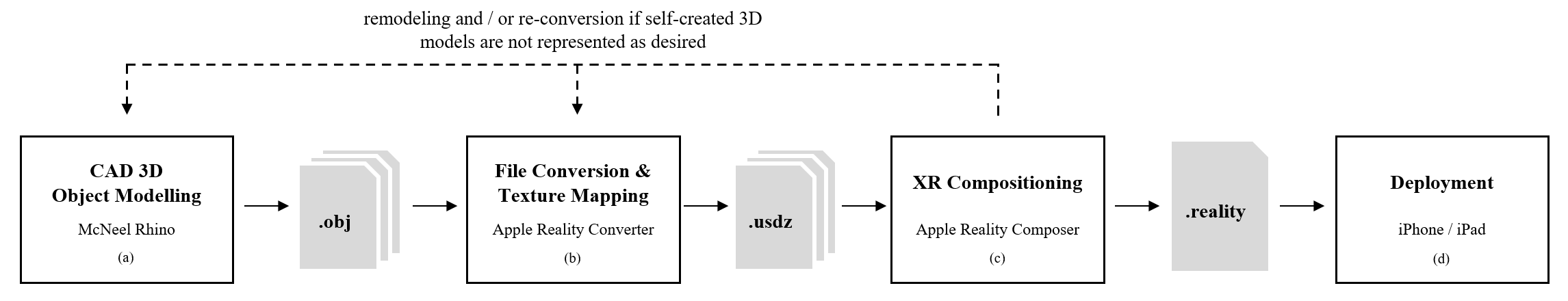}  \\
	%\hline
\end{tabular} 
	\caption{Development process of the MR applications}  
	\label{Software_Process}
\end{figure*}

For creating a MR lecture application, four steps are necessary: create a 3D object geometry; convert file and conduct texture mapping; XR composition; deployment. The development process is shown schematically in Figure \ref{Software_Process}.

The starting point is the creation of 3D models of the components to be visualized based on 2D drawings from the conventional exercise documents (Fig.~\ref{Software_Process} a). In this study, it was performed using the commercial CAD software \textit{McNeel Rhinoceros}. During this step, additional information to be inserted, such as physical dimensions or loads, had also to be modeled in 3D. Subsequently, the 3D geometries were exported as .obj-files, resulting in the loss of any texture information. Material properties are lost in the process and the entire geometry within an .obj-file is later considered as one coherent object. Therefore, the entire frame cannot be exported as one object, but it must be divided into individual components, since, for example, the constructive details are to be highlighted in color.
Before the 3D geometries can be imported into the Reality Composer, they have to be converted into the .usdz-format (Fig.~\ref{Software_Process} b). Apple provides the application \textit{Reality Converter} for this purpose, with which common 3D formats can be converted into .usdz-files. Thereby it is possible to again define the texture mapping for each object, so that this information is retained throughout the rest of the process.

The next step is the composition of the MR content in the Reality Composer (Fig.~\ref{Software_Process} c). For the creation of MR content, it is possible to choose from a selection of predefined elements such as cuboids, arrows and text elements, or to import one's own geometries as .usdz files. The position, size, color and material of the predefined elements can be customized as desired. For imported objects, only the position in space and the size can be adjusted, but not the color. Thus, if a component is to be highlighted after a certain user interaction, two instances of it must be imported and positioned with different texture maps, i.e. colors. For easy handling of the large number of objects in the creation of MR applications, the verification process was divided into several \textit{scenes}, each representing specific content. Thus, in Application A, each substep (i.e. geometry of the component, the acting loads, ...) was represented in a single scene. In application B, the overall system was created in one scene and the design of the structural details was created in separate scenes.
Interaction with the MR content was implemented using \textit{Behaviours}. They consist of two parts: (i) Defining the trigger: this can be, for example, tapping on a 3D object. (ii) Defining the actions to be performed: Here, one can choose from a variety of actions such as scaling, rotating, or showing/hiding objects. These behaviors were used, for example, to display loads that were initially hidden by tapping the corresponding button. Behaviours are also used to switch between scenes in the MR applications. 

The finalized MR compositions were exported as .reality-files, placed on cloud storage, and QR codes with links to them were created (see Fig.~\ref{fig:QRCodes_AppsVids}). These QR codes can be printed on worksheets or inserted in the presentation slides during the exercises so that students can launch the MR applications on their devices conveniently (Fig.~\ref{Software_Process} d).

\subsection{Application Interface and Interaction Design}
\label{subsec:Interface}

\begin{figure} [htbp]
	\centering
	\begin{tabular}{c}
	\hline  
	\cellcolor{hellgrau} \small Sequence Diagram of the developed MR Workflow \\ 
	\hline  
	\vspace{.5mm}
	\includegraphics[width=0.9\linewidth]{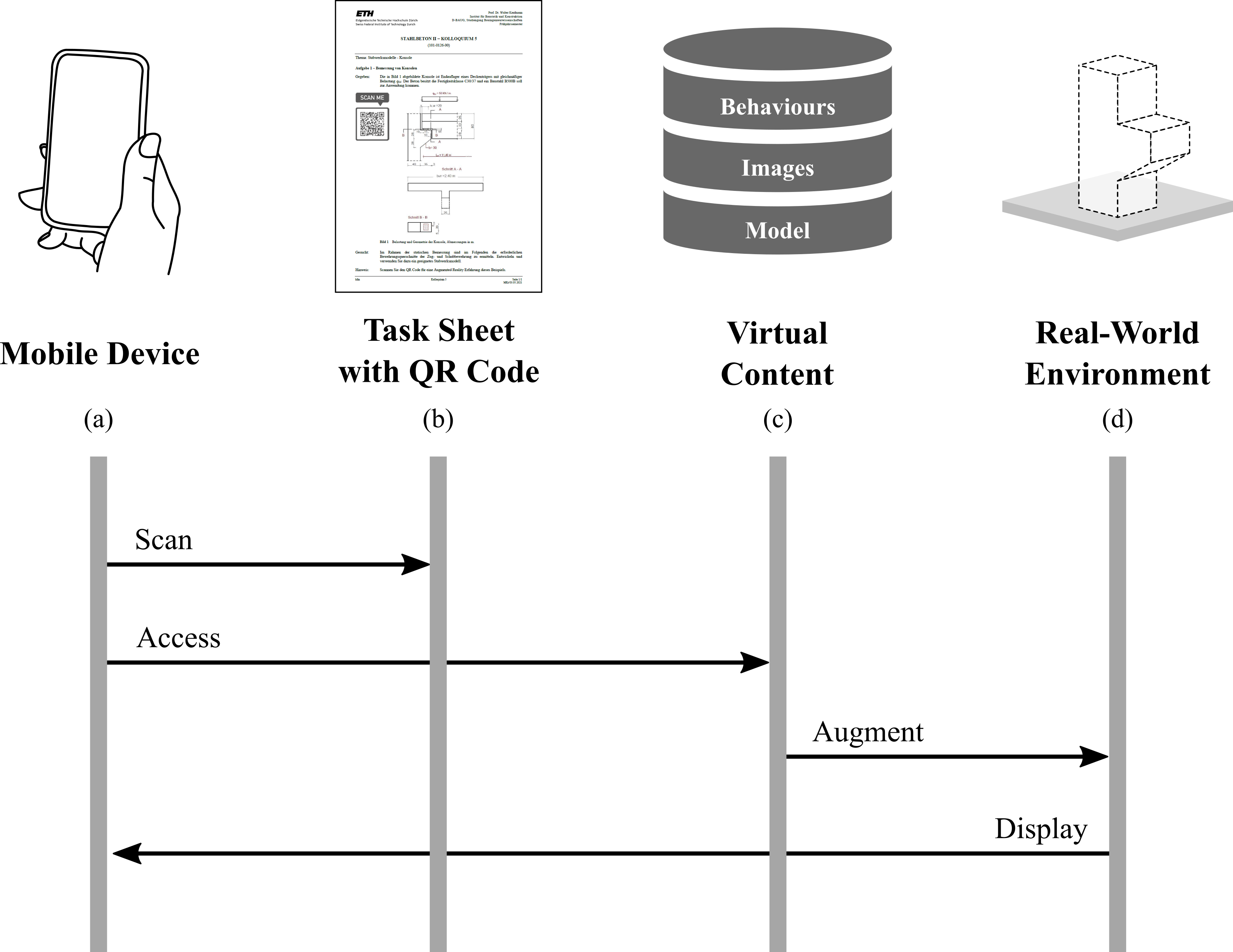}  \\
	\hline
\end{tabular} 
	\caption{Sequence Diagram of the developed MR Workflow}  
	\label{fig:SequenceDiagrammMRApp}
\end{figure}

\begin{figure} [h!]
	\centering
	\begin{tabular}{c}
	\hline  
	\cellcolor{hellgrau} \small Screenshot taken of Concrete Corbel MR App \\
	\hline  
	\vspace{.5mm}
	\includegraphics[width=0.95\linewidth]{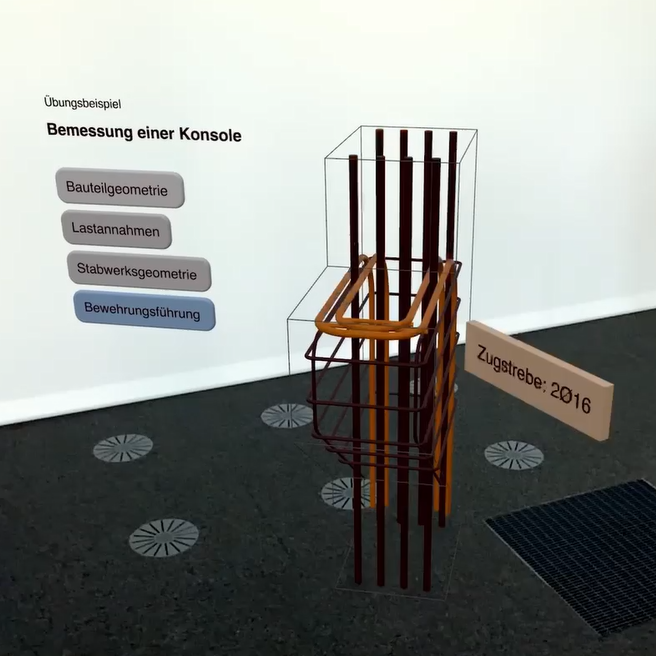}  \\
	\hline
\end{tabular} 
	\caption{Screenshot of Application A showing the rebar of the concrete corbel}  
	\label{ScreenshotARebar}
\end{figure}

\begin{figure*} [h!]
	\centering
	\begin{tabular}{c}
	\hline  
	\cellcolor{hellgrau} \small Screenshot taken of Steel Frame MR App \\
	\hline
	\vspace{.5mm}
	\includegraphics[width=0.65\linewidth]{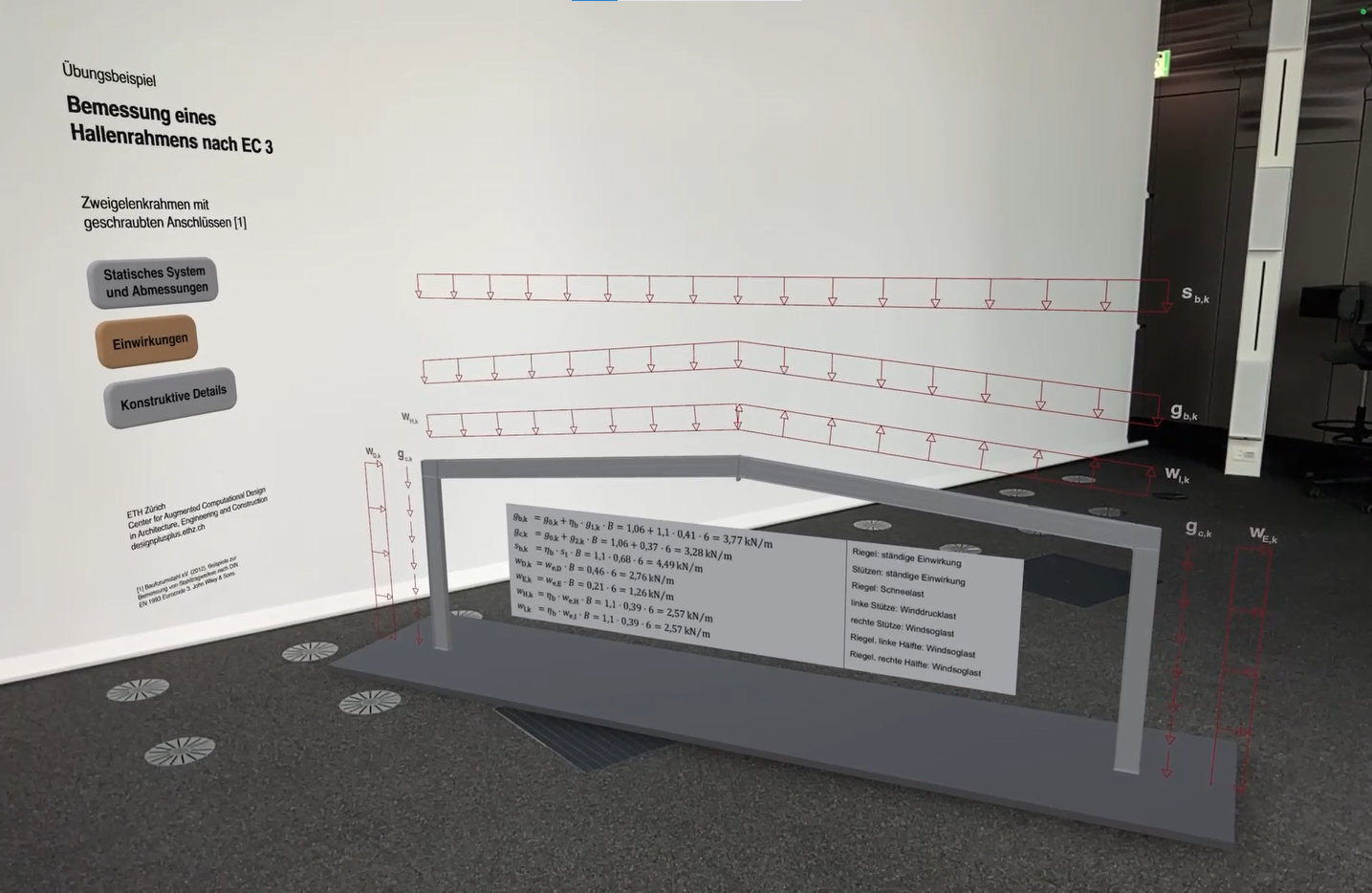}  \\
	\hline
\end{tabular} 
	\caption{Screenshot of Application B showing the whole frame and the acting loads}  
	\label{ScreenshotBFrame}
\end{figure*}

The interactions of the users with the \emph{Struct-MRT} MR app is shown as a sequence diagram in Fig.~\ref{fig:SequenceDiagrammMRApp}. During a lecture, students may access the additional content of a task sheet by using their mobile device (iPhone / iPad) to scan a quick response (QR) code via the built-in cameras. Next, as they open the MR app on their handheld devices, 3D computer generated content including multimedia (e.g. videos, sounds, and images) are displayed in an AR manner (cf. Fig.~\ref{ScreenshotARebar} and Fig.~\ref{ScreenshotBFrame}). 

The MR views of the developed applications are laid out in full-screen mode, supporting both landscape and portrait orientation. Additional graphical interaction widgets are placed along the 3D content depending on the specifics of the deployment example. With this layout, the major portion of the view is kept clear of clutter, often caused by widgets and other elements.

Fig.~\ref{ScreenshotARebar} shows a screenshot of application A. The console placed on the right side in space is complemented by a menu placed on the left side in space. All interactions are done either by tapping on the virtual buttons or on individual component parts. Since no markers are used, the content can be arranged anywhere in the room on a horizontal surface and finger gestures can be used to adjust the position and size. Initially, the console is displayed at 1:1 scale and positioned so that the user can comfortably explore it from a standing position.

Fig.~\ref{ScreenshotBFrame} shows a screenshot of application B. In this figure, the complete steel frame can be seen as a global system upon startup, which is initially displayed at a scale of 1:100 to fit in an ordinary classroom. Figure \ref{ScreenshotBFrame} also shows that extensive calculations from the design documents are displayed via images positioned in space.

\begin{figure*} [htbp]
	\centering
	\begin{tabular}{c}
	\hline  
	\cellcolor{hellgrau} \small Screenshot of Details taken of Steel Frame MR App \\ 
	\hline  
	\vspace{.5mm}
	\includegraphics[width=0.65\linewidth]{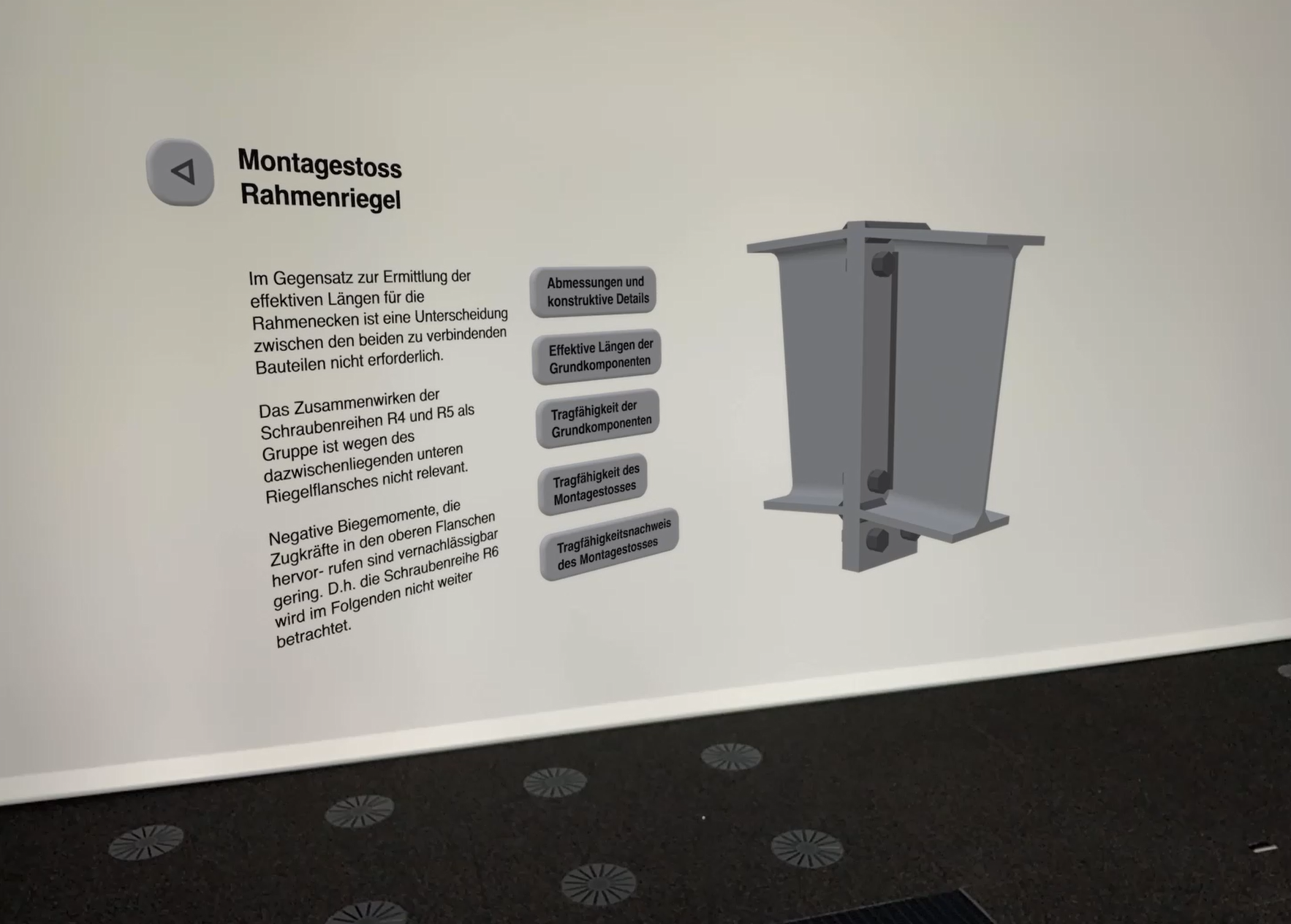}  \\
	\hline
\end{tabular} 
	\caption{Screenshot of Application B showing the explanatory instructions for designing the apex joints of the steel frame}  
	\label{ScreenshotBRoofJoint}
\end{figure*}

By tapping on the structural details in the representation of the entire frame, the user is taken to the separate view (cf. Fig.~\ref{ScreenshotBRoofJoint}). In this, the components are visualized at a scale of 1:1. Again, a spatial menu is used to let the user navigate interactively through the verification process. Text elements and screenshots from the documents are used to show the verification calculations. Depending on the particular process step, additional information such as dimensions is displayed or individual components (e.g. bolts) are highlighted in color.  

In this setting, students may also work collaboratively in groups to discuss aspects of the displayed information. The simultaneous use of multiple devices in a group further enhances participation and encourages interaction between the group members. Furthermore lecturers are enabled to form teams of students and easily implement the tool in the classroom by asking students to use their own mobile devices at no additional cost.

% Design and Assessment of a Mobile Augmented
Acc. to \citep{cuendet2013designing} a good educational system design contains seven key principles: (1) interaction, (2) empowerment, (3) awareness, (4) flexibility, (5) accessibility, (6) immediacy, and (7) minimalism. The authors ensured, that \emph{Struct-MRT} incorporates all these principles to enhance quality of the pedagogical approach. Specifically, utilizing the context-aware mobile MR application to show additional visual content to the exercise sheet along with the instructor’s knowledge on the topic gives empowerment (2) and awareness (3). In addition, the ability to utilize the MR apps individually or in a group setting allows for interaction (1) and flexibility (4), by enabling students and lecturers to cooperate for adapting to various degrees of information and knowledge within a group or between the groups. Concerning accessibility (5) and immediacy (6), students can promptly get to the MR learning materials anyplace and anytime, and receive instantaneous response from the MR tool (after they downloaded the MR applications). Finally, minimalism (7) in both, the visual perception components of the interface and the quantity of accessible functionalities, was reliably observed in the framework. In summary, the research reported in this paper covers all seven key principles and thus provides an integrated learning environment approach for instructors, lecture material, MR devices and students.

\section{Deployment Study: Assessment of the Mixed Reality Applications}
\label{sec:Study}
Having developed and designed the overall technical resp. pedagogical framework and the necessary implementation steps, this section is concerned with the evaluation of the technical performance as well as the deployment amongst lecturers and students. The technical performance was evaluated by the authors, while the methodology was examined trough surveying a test panel after experiencing with the newly developed MR apps.

\subsection{Technical Performance Assessment}
\label{subsec:PerformanceAssessment}

This study employed devices acc. to Table~\ref{tab:DevicesTable_Results} for testing the developed MR applications. For the technical performance assessment the authors investigated the performance indicators (i) average loading time (ALT) and (ii) mapping success, as suggested by \citep{Nowacki2020,Cervenak2019}. A proper "understanding" of the surrounding space by the MR application allows for well mapped surfaces and thus for realistic placement of virtual objects in the real space. The surface detection quality itself determines the user acceptance and experience of the entire application. If accuracy is too low, virtual content will move or appear in places where it would be impossible in the real world, spoiling the general impression and the comfort of using the application. In addition to accuracy, the time needed for loading the app but also surface detection is important as too long time spans for these tasks would weaken the functionality of such applications and again comfort of use. To that end, the MR applications were also tested in various surroundings (office, Immersive Design Lab (IDL), student's apartment (APT)) and light conditions (daylight / DL, artificial light / ATL). For this study, the authors just noted down whether or not the mapping of the MR content in a specific condition was successful. The MR application file size for (A) concrete corbel amounted to 8.2 MB and for (B) steel frame to 35.8 MB.

\begin{table*}[]
\caption{Technical Performance Assessment Overview: Devices, Technical Comparison and Test Results}
\label{tab:DevicesTable_Results}
\resizebox{\textwidth}{!}{%
\begin{tabular}{|l|c|c|c|c|c|c|c|c|c|c|c|c|c|c|c|c|c|}
\hline
\multirow{4}{*}{\textbf{Device}} &
  \multirow{4}{*}{\textbf{Processor}} &
  \multirow{4}{*}{\textbf{Cores}} &
  \multirow{4}{*}{\textbf{\begin{tabular}[c]{@{}c@{}}RAM\\ {[}GB{]}\end{tabular}}} &  \multicolumn{7}{c|}{\textbf{A: Concrete Corbel}} &
  \multicolumn{7}{c|}{\textbf{B: Steel Frame}} \\ \cline{5-18} 
 &
   &
   &
   &
  \multirow{3}{*}{\textbf{\begin{tabular}[c]{@{}c@{}}ALT\\ {[}s{]}\end{tabular}}} &  \multicolumn{6}{c|}{\textbf{Mapping Success}} &
  \multirow{3}{*}{\textbf{\begin{tabular}[c]{@{}c@{}}ALT\\ {[}s{]}\end{tabular}}} &  \multicolumn{6}{c|}{\textbf{Mapping Success}} \\ \cline{6-11} \cline{13-18} 
 &
   &
   &
   &
   &
  \multicolumn{2}{c|}{\textbf{Office}} &
  \multicolumn{2}{c|}{\textbf{IDL}} &
  \multicolumn{2}{c|}{\textbf{APT}} &
   &
  \multicolumn{2}{c|}{\textbf{Office}} &
  \multicolumn{2}{c|}{\textbf{IDL}} &
  \multicolumn{2}{c|}{\textbf{APT}} \\ \cline{6-11} \cline{13-18} 
 &
   &
   &
   &
   &
  \textbf{DL} &
  \textbf{AL} &
  \textbf{DL} &
  \textbf{AL} &
  \textbf{DL} &
  \textbf{AL} &
   &
  \textbf{DL} &
  \textbf{AL} &
  \textbf{DL} &
  \textbf{AL} &
  \textbf{DL} &
  \textbf{AL} \\ \hline
iPhone 8              & A11 Bionic  & 6 & 2 & 7 & yes & yes & yes & yes & yes & yes & 20 & yes & yes & yes & yes & yes & yes \\ \hline
iPhone 8 Plus         & A11 Bionic  & 6 & 3 & 6 & yes & yes & yes & yes & yes & yes & 16 & yes & yes & yes & yes & yes & yes \\ \hline
iPhone X              & A11 Bionic  & 6 & 3 & 6 & yes & yes & yes & yes & yes & yes &  16  & yes & yes & yes & yes & yes & yes \\ \hline
iPad Pro (2017)        & A10X Fusion & 6 & 4 &  6 & yes & yes & yes & yes & yes & yes &  16  & yes & yes & yes & yes & yes & yes \\ \hline
iPad Pro (4th gen.) & A12Z Bionic      & 4 & 6 & 6  & yes & yes & yes & yes & yes & yes &  13  & yes & yes & yes & yes & yes & yes \\ \hline
iPad Air (3rd gen.) & A12 Bionic    & 6 & 3 & 5 & yes & yes & yes & yes & yes & yes & 15   & yes & yes & yes & yes & yes & yes \\ \hline
\end{tabular}%
}
%\vspace{5mm}
{\tiny ALT: average loading time \hspace{5mm} DL / AL: daylight / artificial light conditions \hspace{5mm} IDL / APT: Immersive Design Lab / Student's Apartment}
\end{table*}

Thanks to the above-mentioned features, it was possible to create a base for tests of both frameworks, and finally presentation of strengths and weaknesses. Apps were tested on several devices (Table 1).

For the technical performance assessment the authors found smooth operation for almost all tested devices. The MR app is robust under almost all diverse typical learning conditions (environments, light conditions).

\subsection{Deployment Study Construction and Conduction}
\label{sec:Study_construction}
After technical performance assessment, the designed MR applications were piloted to different student and instructor groups from architecture and civil engineering, who were provided with the MR apps in two stages: (i) close colleagues of the authors (experimental group); (ii) a panel of students and instructors of civil engineering and architecture from ETH Zurich. The deployment study investigated the following objectives:
\begin{itemize}
\item Determination of user experience satisfaction, ease-of-use and interaction.
\item Determination of how the participants view the use of MR in engineering education for knowledge transfer. Thereby, their previous experience with extended reality applications was also taken into account.
\item Identification of possible deficits and potentials for improvement of the MR apps. Both the visual and contextual conception as well as the technical implementation of the MR experiences on the end devices should be considered.
\item Identification of further suitable use cases in the wide-ranging subject areas of civil engineering education.
\end{itemize}
Since the developed MR applications are prototypes for the purpose of demonstration, a full-scale deployment in a civil engineering course was refrained from. The experimental group was monitored by the authors during the test phases without interaction or support (as gaining an impression of MR app usage stood in the foreground) and subsequently also completed a survey investigating the participants' attitudes toward learning structural analysis in an MR environment. The MR applications were then made available online to students and instructors of civil engineering and architecture for independent examination. As part of this, they were asked for their opinion regarding MR in civil engineering education by means of a survey. The survey was adapted from \citep{Turkan2017} as well as \citep{wojciechowski2013a} and was based on a Technology Acceptance Model \citep{davis1989a}.The results of the study were evaluated descriptively and additionally differences in assessments between various subgroups of participants were statistically analyzed for significance.

To conduct the study, a Google Form was set up to be used for both participant instruction and data collection. The study consisted of three phases, which were completed independently by each participant:

The first step is a textual introduction to the topic accompanied by questions about demographics and previous use of XR applications. It is also asked whether the participant owns an iPhone / iPad in order to be able to test the applications on their own end device.

In the second phase, the applications are examined by the participants. Those participants who own an iPhone / iPad can retrieve applications from a cloud repository and explore them natively on their end device. They will not be given any instructions on how to use applications. Participants who do not own an iPhone or iPad had the opportunity to watch a video demonstration of the applications. For this purpose, the applications were operated in advance by the authors in an exemplary manner and the screen was recorded during this process. These participants therefore watch videos without being able to interact with the applications themselves.

In the third phase, the participants are asked a number of questions. Firstly, their attitude towards the use of MR in the design of components is surveyed. For this purpose, they are asked questions based on a Technology Acceptance Model \citep{davis1989a}. These questions are aimed at the participants' assessment of the interface style, perceived usefulness, perceived enjoyment, perceived ease of use, attitude toward using, and intention to use by means of a 5-point Likert scale with ratings from (1) \textit{strongly disagree} to (5) \textit{strongly agree}. Second, participants are asked to indicate in which other subfields of civil engineering they find the use of MR useful respectively most useful. Participants were able to choose from structural engineering, construction management, geotechnical engineering, hydraulic engineering, and infrastructure management or name other areas. Finally, participants are asked to indicate what functionalities an MR application should have for teaching civil engineering. Again, they were given some choices (\textit{enable parametric representations}, \textit{represent manufacturing steps of the component}, ...), and an optional text box was provided to collect additional functions requested by the participants.

\begin{figure} [H]
	\centering
	\begin{tabular}{c}
	\vspace{.5mm}
	\includegraphics[width=0.75\linewidth]{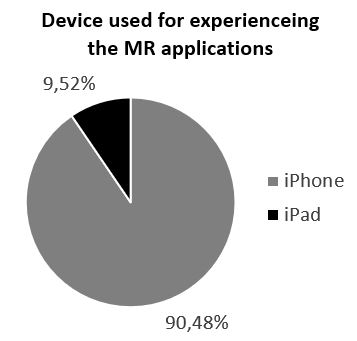}  \\
	%\hline
    \end{tabular} 
	\caption{Breakdown of used Apple devices, $N = 42$.}
	\label{DeviceUse}
\end{figure}

\subsection{Participants}
\label{sec:Study_participants}
The study was conducted amongst mostly civil engineering and architecture students in bachelor, master and PhD program as well as instructors from the civil engineering department of ETH Zurich. An invitation email was distributed by the authors through departmental mailing lists. To encourage participants, the purpose of developing MR application to support students' learning in structural engineering lectures was outlined. After two weeks, a total of 89 responses were received. The panel splits in 48 persons with civil engineering background, 27 with architectural background and 14 with an other educational / professional background. Another split shows, that 38 people have a Bachelors degree, 30 a Masters degree, 5 a PhD and the rest a high school or other degrees. Further demographic information on age, gender or residency were not considered at this stage.

The survey highlighted, that 42 participants (almost half of the panel) possess Apple devices and hence could access the MR apps, where roughly 10\% of of these students used the iPad. The number of people from the panel owning a smartphone or tablet device with other operating systems is seen similarly high as technology is expected to be embedded and playing an important role in students’ daily lives. The other half of the panel without Apple devices was provided with prerecorded videos of the usage and functionality of the MR apps, so that hypothesis testing between the group with the MR experience and the group informed by the video can be established.

\subsection{Pilot Study Results and Discussion of Findings}
\label{sec:Study_findings}
The next subsections introduce the quantitative and qualitative results of the survey and subsequent processing, a discussed is given in Sec.~\ref{sec:Conclusions}.

The comparison of results uses $p$-value tests, which calculates the probability of obtaining a test statistic result at least as extreme as the one that was actually observed, with the assumption that the null hypothesis $H_0$ is true \citep{mendenhall2016statistics}. Specifically the authors conduct 2-sample t-tests at 5\% significance level, where the null hypotheses are specified in the subsequent sections.

\begin{figure} [H]
	\centering
	\begin{tabular}{c}
	\vspace{.5mm}
	\includegraphics[width=0.99\linewidth]{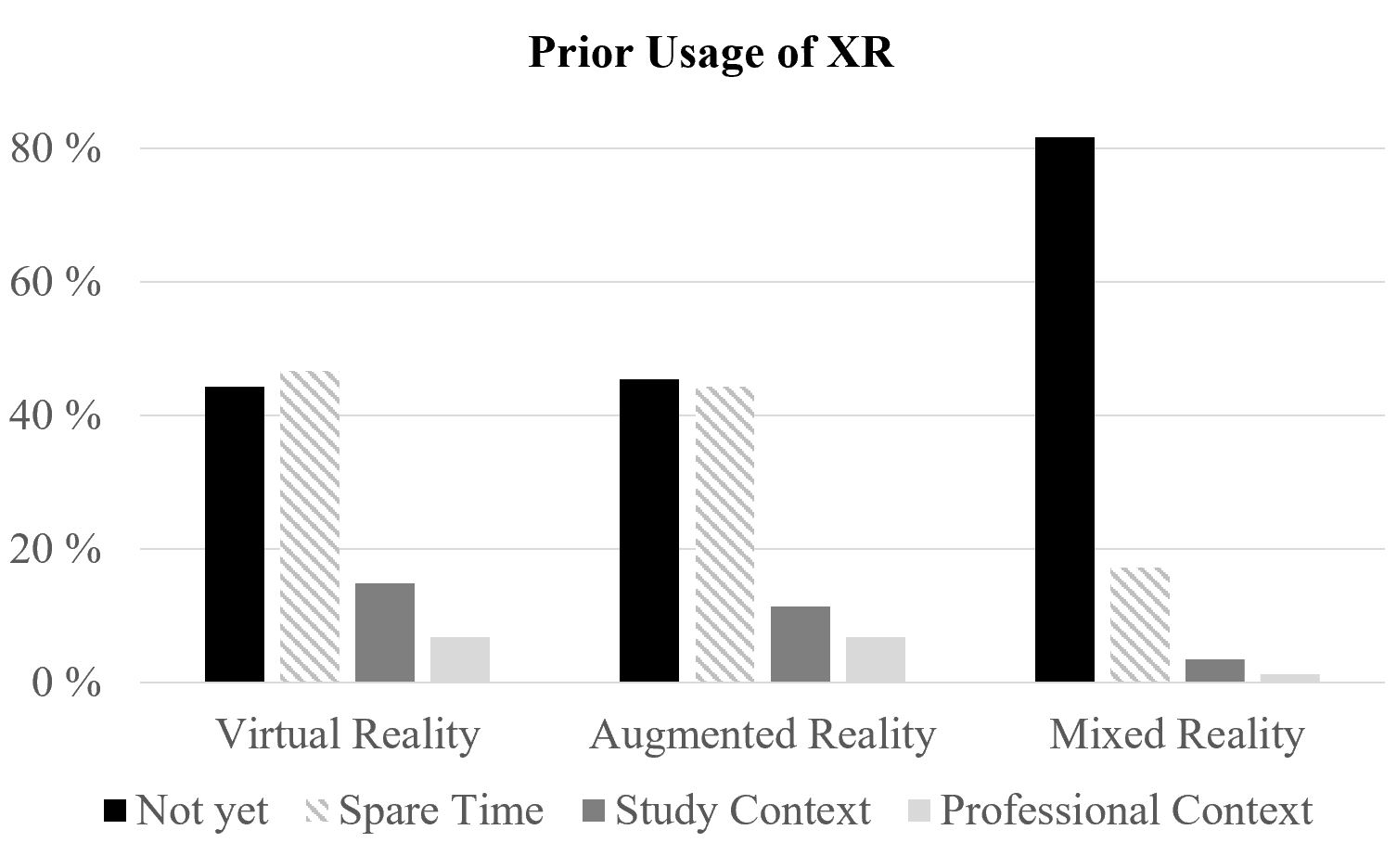}  \\
	%\hline
    \end{tabular} 
	\caption{Breakdown of participants depending on previous use of XR, divided into VR, AR, and MR as well as the area in which these technologies were used.}
	\label{PriorUsageXR}
\end{figure}

\subsubsection{Prior Usage of XR}
\label{sec:Study_findings_priorUsage}
The obtained answers w.r.t. prior usage of XR by the panel are given in Fig.~\ref{PriorUsageXR}. On the one hand, circa half of the panel did not yet experience AR or VR, where in contrast almost the same fraction of persons did use either VR or AR in their spare time. Interestingly, 82\% of persons indicate no MR experience at all or just 17\% in their space-time. This is again seen to stem from the fact, that the vast majority of persons are not familiar with the distinct definitions of AR/VR/MR (cf. Sec.~\ref{sec:intro} resp. Sec.~\ref{sec:Literature_SoA_XR}). In addition, the absence of XR technology in teaching and engineering practice becomes obvious given the low numbers of recognition from the panel. A deeper insight into XR awareness is not possible given the data as no question about prior perception of XR technology in whatsoever form was asked.

\begin{table*}[p]
    \caption{Participants' attitudes toward building with subdivision into subgroups to evaluate differences in opinion for significance (at significance level of $\alpha$ = 0.05)}
    \centering
    \begin{tabular}{c}
    \includegraphics[width=0.70\textwidth]{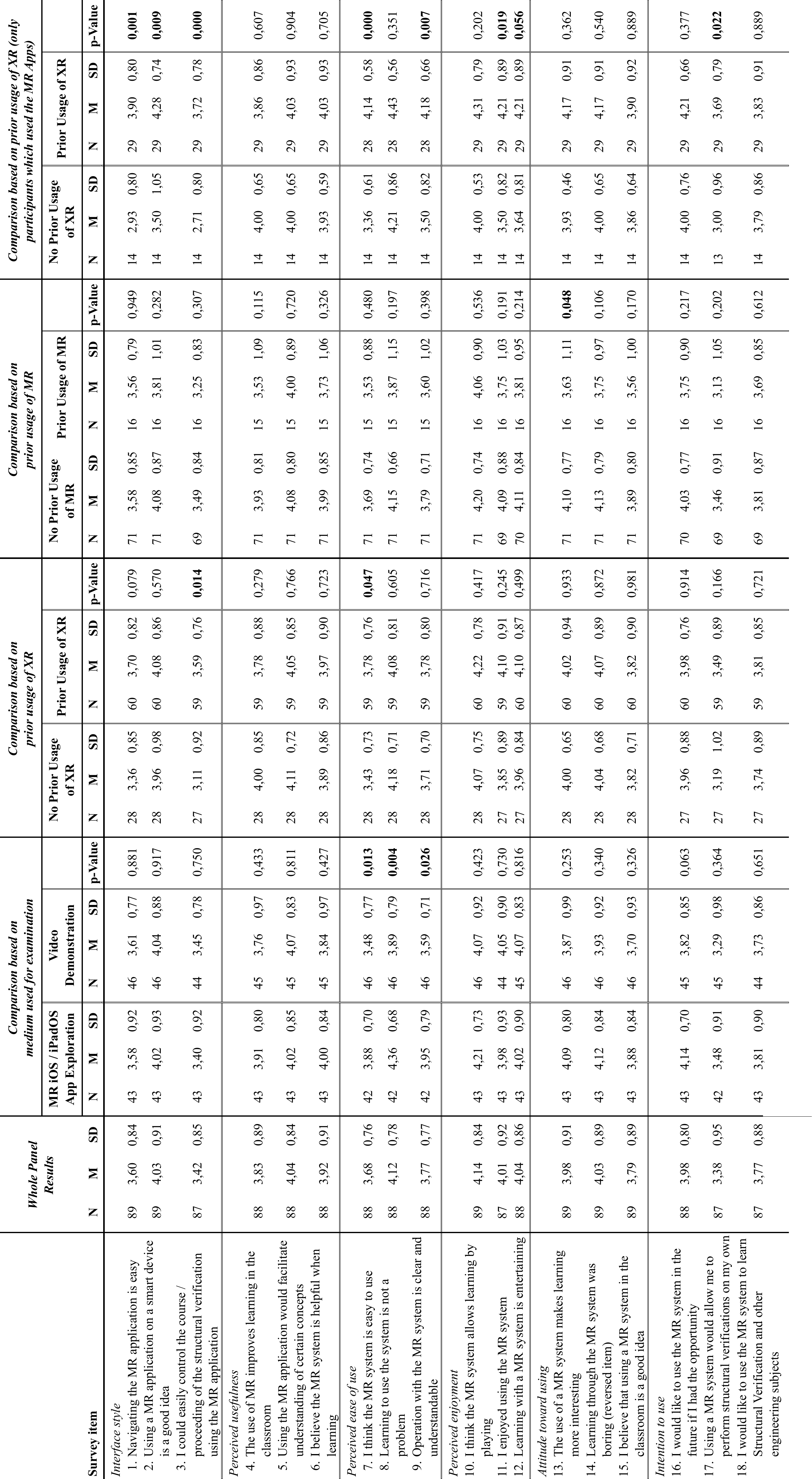}
    \end{tabular}
    \label{tab:AttitudeXRApps}
\end{table*}

\subsubsection{Participants Attitude towards using XR for explaining Structural Design and Verification}
\label{sec:Study_findings_perspectivesVerification}

The obtained results of the survey are summarized in Tab.~\ref{tab:AttitudeXRApps}, where $N$ is the number of respective answers, "M" resp. "SD" is the mean and standard deviation of the rating on the Likert scale. A couple of observations can be made in addition to the testing of several hypotheses H1 - H5, which are detailed in the following:
\begin{itemize}
\item H1: The advantages of Mixed Reality are easy to grasp for students
\item H2: It is easy for students to learn to work with Mixed Reality technologies and tools.
\item H3: Students realize the advantages of Mixed Reality technologies for teaching and lecturing.
\item H4: Students enjoy working with Mixed Reality technologies and tools
\end{itemize}

Computing the mean values of the ratings of the whole panel acc. to Tab.~\ref{tab:AttitudeXRApps} for
\begin{itemize}
    \item questions 1,3 and 7-9 delivers $M = 3.72$ and hence there is evidence for the hypothesis H1 being true.
    \item questions 7-9 delivers $M = 3.86$ and hence there is evidence for the hypothesis H2 being true.
    \item questions 4-6, 13-18 delivers $M = 3.86$ and hence there is evidence for the hypothesis H3 being true.
    \item questions 10-12 delivers $M = 4.06$ and hence there is evidence for the hypothesis H4 being true.
\end{itemize}

One core assumption of using MR for teaching structural design and verification of the authors was, that displaying context-related 3D models to show the content is a good solution and the visualization features were considered to be highly beneficial for learning. Hypothesis H4 in particular supports, that the panel enjoyed representation of the structures in an easily understandable MR way as opposed to 2D images typically found in textbooks or exercise sheets. From these observations, it can also be verified, that a great majority of students can be seen as visual learners and there is a general acceptance for XR methods in teaching and eventually professional life. This finding is supported by other studies, where in addition to verifying most engineering students being visual learners, it is found, that they learn better in interactive teaching environments \citep{shirazi2015design, dong2013a}.

\citep{Turkan2017} specifically address their results of question 14 with a low median rating value together with a high standard deviation and concluded, that some students merely played with the application features instead of working on the problem as visualization features may have overwhelmed students. In our study, the rating is even higher and hence would lead to the conclusion, that the panel at hand did either play even more or was even more overwhelmed. However, the authors do not agree on this conclusion but deduce, that immersive interaction with the displayed content encouraged the panel to intensively interact with the content without loosing interest quickly. In addition, this finding together with correctness of hypotheses H2 and H4 can be seen as a proof of good application design.

Further statistical evaluations of the panel are conducted for the survey results comparison based on the
\begin{itemize}
    \item H5: medium used for examination.
    \item H6: prior usage of XR.
    \item H7: prior usage of MR.
    \item H8: prior usage of XR where only participants which used the MR Apps are considered.
\end{itemize}
The statistical evaluations for a 2-sample t-test of hypotheses H5-H8 are given in Tab.~\ref{tab:AttitudeXRApps}. For H5, only the question block on perceived ease of use (7-9) is statistically significantly different. Comparing the mean ratings of participants using iOS devices to those exposed to the video, it is clear, that using the mobile device for inspection of the MR application delivers more insight into the ease of use than watching a video. For H6 questions 3 and 7 show statistically significant differences for the panel ratings, where it is interesting, that those two questions are concerned with the usage of the application. From the point, that this hypothesis test wants to quantify the influence of prior XR usage, it is clear, that people already familiar with XR environments can naturally handle and navigate more easily compared to novices in that field. Testing of hypothesis H7 delivers only a significant influence of prior MR usage on question 13 concerned with the motivational aspect of learning by MR. Testing of H8 finally is interesting, as several influences of prior XR use can be found amongst the participants having actually used mobile devices to experience \emph{Struct-MRT}. These people rated the interface style (questions 1-3) and ease of use (questions 7,9) significantly higher and additionally indicating greater enjoyment during use (questions 11,12). Furthermore they felt enabled to perform structural verifications on their own (questions 17) by the \emph{Struct-MRT} tool. These findings highlight the significant influence of XR prior knowledge and underlines motivational aspect for the learning process (understand on their own in an accessible way).

\subsubsection{Participants Perspectives on the Use of XR for Civil Engineering Education}
\label{sec:Study_findings_perspectivesEducation}
\begin{figure} [H]
	\centering
	\begin{tabular}{c}
	\vspace{.5mm}
	\includegraphics[width=0.98\linewidth]{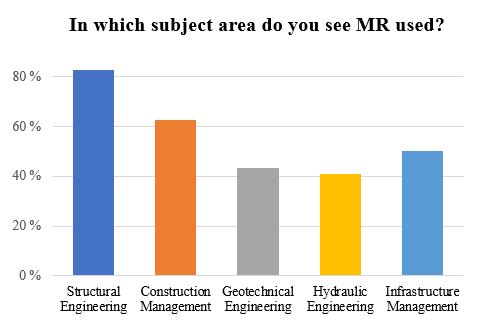}  \\
	%\hline
    \end{tabular} 
	\caption{MR usage in civil engineering subfields}
	\label{fig:RelevantAreas_1}
\end{figure}

\begin{figure} [H]
	\centering
	\begin{tabular}{c}
	\vspace{.5mm}
	\includegraphics[width=0.98\linewidth]{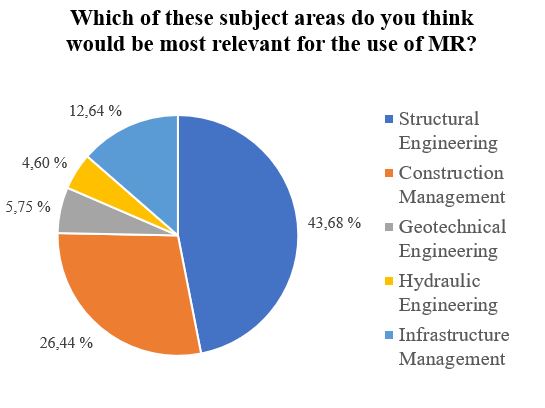}  \\
	%\hline
    \end{tabular} 
	\caption{Most relevant civil engineering subfields for usage of MR}
	\label{fig:RelevantAreas_2}
\end{figure}

The last block of the survey concentrated on participants view on relevant areas of civil engineering education for usage of MR applications. The distribution of choices from different subfields is displayed in Fig.~\ref{fig:RelevantAreas_1} and Fig.~\ref{fig:RelevantAreas_2}.

The panel clearly sees the most application cases of MR in structural engineering, followed by construction resp. infrastructure management. Some use of MR is seen in geotechnical and hydraulic engineering. The results indicate, that perception and understanding of especially complex structural systems with elaborated design and verification tasks can greatly be supported by MR. On the other hand, the authors want to emphasize, that the results of Fig.~\ref{fig:RelevantAreas_1} have to be seen in the light of an unquantifiable bias towards structural engineering as the two examples shown to the panel for the survey did not cover geotechnical or hydraulic engineering tasks, and potentially more people with structural engineering interests were motivated to take the survey (as the authors are mainly active in this field).

%\begin{figure*} [b]
%	\centering
%	\begin{tabular}{c c}
%	\hline  
%	\cellcolor{hellgrau} \small (a) Relevant areas for usage of MR & \cellcolor{hellgrau} \small (b) Most relevant area for usage of MR \\
%	\hline  
%	\vspace{.5mm}
%     \includegraphics[width=0.47\linewidth]{Paper_1/Diagrams/Diagram_SubjectAreas.jpg} &
%     \includegraphics[width=0.47\linewidth]{Paper_1/Diagrams/Diagram_MostRelevantSubjectArea.jpg} \\
%	\hline
%    \end{tabular} 
%	\caption{Participants perspectives on relevant areas of civil engineering education for usage of MR application}  
%	\label{fig:RelevantAreas}
%\end{figure*}

\begin{figure*}
	\centering
	\begin{tabular}{c}
	\vspace{.5mm}
	\includegraphics[width=0.8\linewidth]{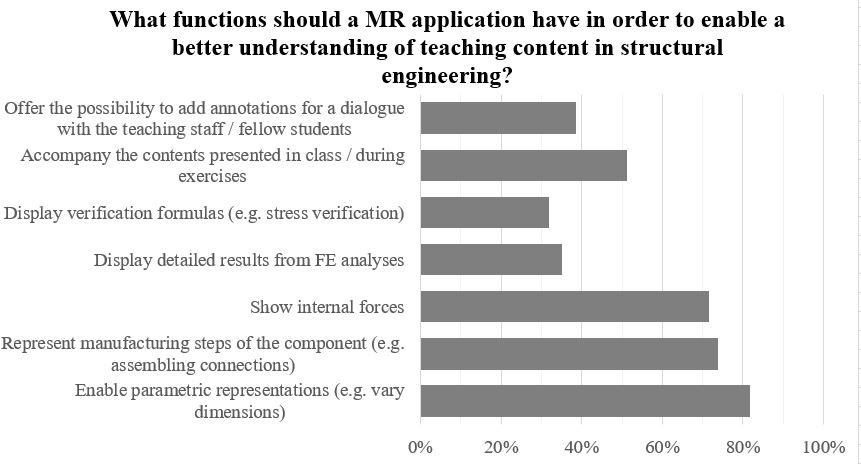}  \\
	%\hline
    \end{tabular} 
	\caption{Requested Functionalities}
	\label{fig:RequestedFunctions}
\end{figure*}

Participants also pointed out some limitations and potential future enhancements of the \emph{Struct-MRT}, cf. Fig.~\ref{fig:RequestedFunctions}. There are three major topical blocks for the requested future features: pedagogical, structural engineering content, and MR interaction. Acc. to the first two questions, ca. half of the panel wishes to see improved MR interaction capabilities in the future. W.r.t. structural engineering content the majority of participants requests displaying internal forces. The most requested future features are concerned with construction tasks (assembling) and gaining further structural insight into the problem through parametric representations. These findings are in accordance with the results shown in Fig.~\ref{fig:RelevantAreas_1} and Fig.~\ref{fig:RelevantAreas_2}, as the main application domains of \emph{Struct-MRT} were seen in structural engineering (design and verification) and construction management (assembly etc.).

Finally, the authors want to pick two exemplary statements (presumably from instructors) and their view on \emph{Struct-MRT}. One participant wrote: \textit{Don't forget that forcing the teacher to actually write equations and diagrams by hand forces the teacher to SLOW DOWN....and it is exactly this slowed pace and the watching of the steps of thought required to think through solving the design problems that helps students learn. PowerPoint is an absolutely terrible medium for all subjects that rely heavily on multi-step math. Perhaps MR might help, but I learned best with the overhead projectors, dry-erase boards, and chalkboards that slowed down my teacher's explanations and forced them to write - like we were writing in our notes - and talk at the same time. Bring back these old analog tools before you invest in fancy video games.} The authors partly agree to the raised point of pace in lecturing, on the other hand this sentence emphasizes, that both, instructors and students, have to get familiar with the XR technology. This raises the need for training of instructors as well as students and a transformation of the established lecturing style with XR augmented tools such as \emph{Struct-MRT}. The authors are convinced, that \emph{Struct-MRT} will not replace a structural engineering lesson but rather support individual learning by communicating complexity in a modern, playful and understandable visual way. Another participant wrote \textit{Incredibly good idea, for years engineering courses (specifically civil engineering) were left behind from the digital revolution. Due to the sheer size of what civengs work on, it is impossible to have nice models for everything, and this creates a gigantic disconnect between theory and reality. This seems the first fundamental step for civil engineering (and related fields) to get up to date and offer user friendly ways to experiment and learn.} Here again, the authors agree, that modern tools will play a more important role in especially the professional life of students and hence introduction of these methods in teaching is absolutely necessary as soon as possible.

These individual statements together with the indications of this section from the survey support the authors to improve the \emph{Struct-MRT} application design in the future, where the majority of limitations so far were due to the pilot study nature as presented here.

\section{Conclusions}
\label{sec:Conclusions}
The objective of the research reported in this paper was to design, implement and assess a context-aware mobile MR tool called \emph{Struct-MRT} to augment lectures teaching design and verification to students of civil/structural engineering. Specifically two example exercise lectures for designing a concrete corbel and a steel frame were enhanced using 3D and multimedia virtual information. These two MR applications were implemented for iOS devices (iPhones / iPads) and subsequently assessed for technical and paedagogical performance. The latter was achieved by conducting a survey amongst students and instructors at ETH Zurich from architecture and civil engineering. The results of this pilot study and their evaluation in summary indicate, that the XR have not yet arrived sufficiently in everyday studying or teaching, yet the picture of a broad acceptance of this technology among civil engineers is emerging and that the \emph{Struct-MRT} tool has significant potential to enhance students’ learning of concepts of structural design and analysis. To that end, \emph{Struct-MRT} provides better learning support capabilities for barrier removal between students and technology in education through the interactive workspace and encouraged collaboration and interaction between students and course contents by immersing participants in a multimedia-enabled learning environment. \emph{Struct-MRT} was rated an interesting, helpful, and motivational approach for gaining more in-depth and long-lasting knowledge while allowing interactive and autonomous studying as well as collaborative practicing with other students without the need for assistance by a lecturer. However, this research at the same time uncovered requested features by the students for future enhancement and improvement of the interface as well as interaction design next to more detailed contents (internal forces, construction steps, etc) of \emph{Struct-MRT}. In addition to the potentially transformative added value of MR in teaching structural design and verification, the comments of the panel simultaneously gave insight into technological, organisational, and cognitive challenges of using MR in teaching and learning. A summary of identified chances, potentials and risks by \emph{Struct-MRT} is given in Tab.~\ref{tab:ChancesRisks}. The following two subchapters specifically elaborate on technical and pedagogical conclusions of this work.

\subsection{Technology Conclusions}
With the current hardware improvements, XR technologies continue to experience significant hype amongs society, associated with technology research and marketing investments on the side of manufacturers and software providers. Despite that the authors successfully elaborated a framework for transforming technical drawings into MR usable objects, the pilot implementation indicated the need for future research to circumvent the cumbersome working processes of file-conversion of the status-quo. In addition, slow loading times or system crashes uncover that more complex technical content is is still challenging to handle for mobile XR devices in currently available XR development environments. While the survey indicated positive feedback on the current implementation and deployment of \emph{Struct-MRT}, future enhancements of the interaction interface and additional content features are requested. Developing and testing these two MR applications has revealed several challenges that make them difficult to use broadly in civil engineering courses: 
\begin{itemize}
\item To the authors' knowledge, there is still no solution that enables intuitive development of MR compositions from modeling to cross-platform deployment. For the use on other devices, the teaching content would have to be reproduced again in dedicated software tools.
\item As can be seen in Figure \ref{Software_Process}, the process from 3D modeling of the components to exporting the final MR composition is very fragmented for this chosen implementation solution. This proves to be a difficulty especially when there is an error in the modeling of a component (e.g., component exported as a whole instead of being divided into individual sub-components) or its texture was not specified correctly (e.g., wrong color assigned in the Reality Converter). Thus, several iterations of exporting, converting and re-importing the 3D geometries into the Reality Composer were necessary in some cases.
\item Important information for the verification like physical dimensions or mathematical formulas cannot be created directly in the Reality Composer, but have to be modeled in the CAD program and then exported individually as 3D geometries.
\item Using .reality-files allows for a native display of the MR content on the iPhones / iPads. However, it does not allow buttons to be fixed in position as controls on the screen. To overcome this, additional 3D elements had to be placed in the space and were made interactive using behaviours. 
\item During several tests by the authors, it became apparent that Application B occasionally crashed when starting the demonstrators process or when switching between scenes. Causes for this could be the large number of individual, detailed modeled objects as well as the concurrent high number of Behaviours, which are necessary to represent the verification in the appropriate scope. The size of the files and the stability of the application could not be directly influenced during development.
\end{itemize}

\begin{table*}
\caption{Opportunities and Risks of using \emph{Struct-MRT}}
\label{tab:ChancesRisks}
\resizebox{\textwidth}{!}{%
\begin{tabular}{|l|l|l|}
\hline
\textbf{} &
  \textbf{Chances / Potential} &
  \textbf{Risks} \\ \hline
\textbf{\begin{tabular}[c]{@{}l@{}}Technology / \\ Media specific\end{tabular}} &
  \begin{tabular}[c]{@{}l@{}}- total control of the learning environment via IT engineering\\ - vividness and tangibility of learning content (simulation)\end{tabular} &
  \begin{tabular}[c]{@{}l@{}}- cost-benefit trade-off\\ - implementation effort\\ - accessibility for domain experts\\ - technological risks from non-maturity status\\ - missing technical standards / vendor lock-in\\ - new interaction concepts not mature or lengthy\end{tabular} \\ \hline
\textbf{\begin{tabular}[c]{@{}l@{}}Learning process / \\ education specific\end{tabular}} &
  \begin{tabular}[c]{@{}l@{}}- MR promotes learning as a situational process\\ - reinforcement of the learning experience through presence\\ - different learning types can be addressed equally and simultaneously\\ - multisensory / multimodal learning\\ - context-sensitive and individualized learning support\\ - statistics on learning scenario resp. learning progress\\ - technology-related motivation can be easily triggered\end{tabular} &
  \begin{tabular}[c]{@{}l@{}}- technology-driven use instead of didactic-driven use\\ - teaching/learning concepts / conceptual didactic basis is missing\\ - media competence of teachers and students currently not (yet) given\end{tabular} \\ \hline
\end{tabular}%
}
\end{table*}

\subsection{Pedagogical / Organisational Conclusions}
The topic of XR is not new, but the survey revealed that instructors and students – presumably not limited to  Civil Engineering – lack knowledge of XR-based scenarios in everyday life but also teaching. Universities hence have to provide step-by-step instructions on how to use XR applications to guide instructors and students alike. Introduction of new technologies is definitely the future of education, hence modernisation of curricula together with establishing awareness amongst faculty is necessary and crucial. Taking advantage of the opportunities is the essential step in order to attract attention of students and lecturers and to establish a comforting habit for the use of XR technology. Especially in design and verification of civil engineering structures, an understanding of the parallel applications of experiments and simulation models provides the cause for XR systems to be beneficial. MR deployments can thus be seen as a fitting method bridging the space between the real environment and simulation models from the virtual environment. Taking into account the results of the survey and their analysis, it can be concluded that \emph{Struct-MRT} has potential for development to an effective pedagogical tool to supplement traditional lecturing using ordinary textbooks and paper exercise sheet by augmenting virtual content at low costs associated with development, implementation and maintenance.  

Combining individual and social constructivism served as a backbone of the research reported in this paper, particularly as \emph{Struct-MRT} allows students not only to work interactively in groups and under supervision of the lecturer, but also to use the tool individually at home to review and reinforce the lecture materials. Learning success is not a question of technology only, but of substantiated planning and implementation of a didactic concept. This enforces interdisciplinary cooperation between different disciplines such as didactics, computer science, psychology and civil engineering in the future for a pedagogically sound development of \emph{Struct-MRT}.

\subsection{Open Questions, Future Research Needs and Outlook}
\label{sec:Outlook}
%\citep{zender2018lehren}

The authors see research needs for the mentioned opportunities and risks summarized in Tab.~\ref{tab:ChancesRisks}, since XR learning applications are still relatively young. At the moment, didactically sound and thus learning success-oriented development of XR teaching methods still shows a rather low degree of maturity in addition to being strongly case-study driven. Especially future research needs to address ethical, physiological, psychological, self-determination, safety and privacy implications affected by the XR technology for use in teaching.

Visual thinking, with specific associations to collaborative learning methodologies and distance learning or virtual learning conditions (such as requested during this Covid19 pandemic), is vital for future lecturing and learning. As outlined in the previous section, cross-disciplinary efforts have to be undertaken to further enhance these methodologies accompanied by performance assessment studies of the value of visual and immersive learning. The findings of this work flow into a concept for the improvement of teaching in structural engineering ETH Zurich.

Future work is concerned with the development of the \emph{Struct-MRT} framework for other operating platforms and includes collaboration among several universities to assess the benefits of \emph{Struct-MRT} in multiple courses using larger and more diverse student and instructor populations. A special focus will be on adding simulation result displaying and interaction capabilities to enable improved interaction with \emph{Struct-MRT} and generative design simulations. It will be interesting to see to what extend the research presented in this paper achieves its final objective of assisting students in gaining longer-lasting visual and conceptual information for structural design and verification tasks. The authors hope that in the long run, the study presented in this paper will direct ways to broaden the scope of its use beyond civil engineering teaching into other science, technology, engineering, and math (STEM) disciplines.

%\begin{figure} [h!]
%	\centering
%	\begin{tabular}{c}
%	\vspace{.5mm}
%	\includegraphics[width=0.95\linewidth]{Figures/DevelopmentProcessHoloDesigner.PNG}  \\
%	%\hline
%   \end{tabular} 
%	\caption{Development Process HoloDesigner, from \citep{Dan2021}}
%	\label{DevelopmentProcessHoloDesigner}
%\end{figure}

\section{Acknowledgments}
\label{sec:Acknowledgments}
The authors would like to acknowledge the campus facilities of the Design++ Initiative of ETH Zurich and especially Dr. Romana Rust for great support and providing devices and software for the developments presented in this paper. Furthermore the contributions of the participants of the survey together with their interest is highly appreciated. Any opinions, findings, conclusions, and recommendations expressed in this paper are those of the authors and do not necessarily reflect the views of all associated members of the Design++ initiative.

\section{Declaration of Interests}
The authors declare that they have no known competing financial interests or personal relationships that could have appeared to influence the work reported in this paper.

%\bibliographystyle{spmpsci}      % mathematics nd physical sciences
	

\bibliographystyle{spbasic}

-------------------------------------------------------
	
\bibliography{PaperText}

\end{document}